\documentclass[a4paper,11pt]{article}
\usepackage{jheppub} 
\usepackage{etoolbox}
\usepackage[T1]{fontenc} 
\usepackage{epsfig}
\usepackage{subcaption}
\usepackage{amsmath, amssymb, slashed, multirow}
\usepackage{color}
\usepackage{float}
\usepackage{slashed}
\usepackage{accents}
\usepackage{ulem}
\def\beq{\begin{equation}}
\def\eeq{\end{equation}}
\def\beqn{\begin{eqnarray}}
\def\eeqn{\end{eqnarray}}

\def\tev{{\rm~TeV}}

\title{Electroweak Dark Matter at Future Hadron Colliders}

\author[1,2]{Tao Han,}
\author[1,3]{Satyanarayan Mukhopadhyay}
\author[1]{and Xing Wang}

\affiliation[1]{PITT-PACC, Department of Physics and Astronomy, University of Pittsburgh, PA 15260, USA}
\affiliation[2]{Department of Physics, Tsinghua University, and Collaborative Innovation Center of Quantum Matter, Beijing, 100086, China}
\affiliation[3]{Department of Theoretical Physics, Indian Association for the Cultivation of Science, Kolkata 700032, India}

\emailAdd{than@pitt.edu}
\emailAdd{tpsnm@iacs.res.in}
\emailAdd{xiw77@pitt.edu}

\preprint{PITT-PACC-1806}
\abstract{In a large class of scenarios, dark matter (DM) particles that belong to a multiplet of the standard model (SM) weak interactions are challenging to probe in direct detection experiments due to loop-suppressed cross-sections. Direct production at colliders is thus crucial to look for such DM candidates, and under current estimates, future runs of the 14-TeV LHC are projected to probe masses of  around $300$~GeV for DM belonging to an ${\rm SU}(2)$ doublet (Higgsino-like), and $900$~GeV for $\rm{SU} (2)$ triplet (wino-like). We examine how far this mass reach can be extended at the proposed 27-TeV high-energy upgrade of the LHC (HE-LHC), and compare the results to the case for a 100-TeV hadron collider. Following a detector setup similar to that of the ATLAS tracking system for the Run-2 LHC upgrade, with a new Insertable B-Layer (IBL),
a disappearing charged track analysis at the HE-LHC can probe Higgsino-like (wino-like) DM mass of up to $600$~GeV ($2.1$~TeV) at the $95\%$ C.L. The monojet and missing transverse momentum search, on the otherhand, has a weaker reach of $490$~GeV ($700$~GeV) at $95\%$ C.L. for the Higgsino-like (wino-like) states. The mass range accessible in the collider searches can be complementary to the indirect detection probes using gamma rays from dwarf-spheroidal galaxies.}

\makeatletter
\patchcmd{\maketitle}{\@fpheader}{}{}{}
\makeatother 

\begin{document}
\maketitle
\flushbottom

\section{Introduction}
\label{sec:intro}
One of the simplest realizations of weakly interacting massive particles (WIMP), that could be a natural candidate for dark matter (DM), is the electrically neutral component of a multiplet of the standard model (SM) weak interaction gauge group ${\rm SU}(2)_{\rm L}$~\cite{strumia_minimal,strumia_minimal2}. The well-known example of such DM candidates is that of wino and Higgsino in the minimal supersymmetric SM (MSSM) with R-parity conservation~\cite{susy_dm_review}. 
For each assignment of the DM spin and the ${\rm SU}(2)_{\rm L}$ quantum numbers, the only free parameter in this model for WIMPs is the DM mass. If we further impose the requirement that a single particle species makes up the entire DM relic density through the mechanism of thermal freeze-out, we arrive at the well-known relation \cite{Lee:1977ua,Goldberg:1983nd,Steigman:2012nb}  
\begin{equation}
\Omega h^2 = 0.11 \left(\frac{2.2 \times 10^{-26}~{\rm cm}^3/{\rm s}} {\langle \sigma_{\rm eff} v \rangle_{{\rm freeze-out}}} \right),
\end{equation}
where, $\sigma_{\rm eff}$ is the effective cross-section, which includes the appropriately Boltzmann-weighted thermal averaged contribution from co-annihilating particles, important for DM belonging to electroweak multiplets~\cite{Griest,Gondolo,Edsjo}. For such electroweak DM candidates, since the annihilation rate is fixed by gauge interactions, in the limit $M_{\rm DM} \gg M_{\rm W,Z}$, it can be expressed by simple relations with only one mass-scale,  $M_{\rm DM}$. For example, for wino-like ${\rm SU}(2)_{\rm L}$ triplets, the effective annihilation rate is approximately given by~\cite{well_tempered}
\begin{equation}
\langle \sigma_{\rm eff} v \rangle_{{\rm freeze-out}} \simeq \frac{3g^4}{16\pi M_{\rm DM}^2},
\end{equation}
leading to the thermal relic abundance of
\begin{equation}
\Omega_{\tilde{W}} h^2 \simeq 0.1 \left ( \frac{M_{\rm DM}}{2.2 {~\rm TeV}} \right)^2.
\end{equation}
The above relation gets modified on taking into account corrections from the non-perturbative Sommerfeld enhancement in wino pair-annihilation~\cite{shigeki_sommerfeld}, and ${\rm SU}(2)_{\rm L}$ triplets of mass around $3$~TeV saturate the observed DM abundance. Similarly, for ${\rm SU}(2)_{\rm L}$ doublet Higgsino-like DM, the corresponding mass scale is around $1$~TeV~\cite{Higgsino_coannihilation}. Thus, one obtains a rather robust prediction for the mass of electroweak DM making up the observed DM density. For DM lighter than the above mass scales the thermal relic density is lower, hence either making them viable candidates for a fraction of the total DM in the Universe, or requiring non-thermal production mechanisms~\cite{Moroi}.
  
The prospects for probing such electroweak DM at underground direct detection experiments depends on the representation of the multiplet under ${\rm SU}(2)_{\rm L}$ and its hypercharge. For Dirac fermions or complex scalars with non-zero hypercharge, tree-level neutral current vector interaction with the $Z$ boson leads to a large spin-independent (SI) scattering rate with nuclei. On the other hand, for Majorana fermions and real scalars, the vector couplings vanish identically. We will focus on the detection prospects for ${\rm SU}(2)_{\rm L}$ triplet and doublet Majorana fermions in this study~\footnote{${\rm SU}(2)_{\rm L}$ doublet fermions with non-zero hypercharge, such as the pure Dirac Higgsino in the MSSM, have a vector interaction with the $Z$ boson. However, effective couplings with the Higgs boson (such as those induced by integrating out the gauginos in the MSSM) generate a small mass splitting, thereby decomposing the neutral Dirac fermion into two Majorana fermions, and avoiding the vector interaction. We also need to ensure that such a mixing with the gauginos does not induce large SI scattering through the Higgs boson exchange. For mass splittings larger than $\Delta m \gtrsim\mathcal{O}(100)$~keV, inelastic up-scattering between two Majorana Higgsino mass eigenstates \cite{dm_inelastic} is also avoided. Such mass splittings are however typically so small that they would not affect the collider analyses in the following sections.}. Their SI interaction rate with nuclei is suppressed, as it is generated only at the one-loop order, and suffer from accidental cancellations between different class of diagrams~\cite{Hisano_DD_Wino,DD_Hill}. For wino-like triplet states with zero hypercharge, the SI cross-section with proton is only mildly sensitive on the DM mass, and in the limit $M_{\rm DM} \gg M_W$ is found to be around $2.3 \times 10^{-47} {\rm cm}^2$, including higher order corrections at next-to-leading order in $\alpha_S$~\cite{Hisano_DD_Wino}. Thus, to probe these DM candidates at the direct detection experiments, we would need multi-ton scale detectors. For Higgsino-like ${\rm SU}(2)_{\rm L}$ doublet Majorana fermions, the rate is further suppressed, and the cross-section is around $10^{-49} {\rm cm}^2$, which is below the irreducible neutrino floor, making a detection challenging.

Indirect detection experiments looking for gamma-ray signals from annihilating DM in low-background dwarf-spheroidal galaxies (dSphs) constitutes a more promising probe for electroweak DM candidates~\cite{shigeki_sommerfeld,strumia_minimal2,wino_indirect_shigeki}. For wino-like DM, there is an enhancement of the annihilation rate in certain mass regions in which the exchange of multiple electroweak gauge bosons between the DM particles in the initial state gives rise to a long-range potential (Sommerfeld enhancement)~\cite{shigeki_sommerfeld}.  The current constraints from the Fermi-LAT search for diffuse gamma ray signal from dSphs excludes wino DM mass below around $400$~GeV, and in a small window around $2$~TeV, if it makes up the whole of DM~\cite{wino_indirect_shigeki,Shigeki_review}. However, for these mass values, thermally produced winos are under-abundant. For this reason, as well as to cover the yet-unexplored window in indirect searches, collider probe becomes necessary. For Higgsino-like DM, the annihilation rate is significantly smaller, and the current dSphs constraints only probe mass values smaller than around $350$~GeV, if they saturate the required DM abundance~\cite{Reece}. Thus for Higgsino-like states, the collider probe is crucial as well.

Directly probing heavy electroweak DM at the 14-TeV LHC is found to be very challenging $-$ primarily because of lower Drell-Yan pair production rates for the heavy DM particle and their charged counterparts, and also for the lack of clean experimental handles which can be utilized to suppress the relevant SM backgrounds. Motivated by the supersymmetric wino and Higgsino scenarios, the ATLAS and CMS collaborations have carried out several different searches for electroweak DM at the LHC~\cite{ATLAS_track_8TeV,wino_13tev_current,higgsino_13tev_current,CMS_8TeV,CMS_13TeV}. 
The main collider signatures in the framework under consideration are governed by the small mass splitting between the charged and neutral components of the EW multiplets of about a few hundred MeV. As such, there are broadly two class of searches which are mostly independent of detailed model assumptions. The first one is the classic monojet and missing transverse momentum search for a pair-produced DM particle in association with a hadronic jet originating from initial state radiation. The second one utilizes the fact that, in the absence of large additional corrections from higher-dimensional operators, the mass splitting between the charged and neutral components of the DM ${\rm SU}(2)_{\rm L}$ multiplets is small $-$ of the order of a few hundred MeV. Thus, the decay length of the electrically charged state is large enough to be observed as a disappearing charged track at the LHC detectors. This additional handle helps reduce the SM backgrounds compared to the first search category with only missing momentum requirements, though it does introduce additional systematic uncertainties in the background estimate. The current LHC lower bound using these search strategies for wino-like (Higgsino-like) states is around 460 (152) GeV at $95\%$ C.L.~\cite{wino_13tev_current, higgsino_13tev_current}, and as we will see later, it is projected to improve to 900 (300) GeV at the end of the high-luminosity LHC run (HL-LHC) for wino (Higgsino) states.

Studies on the prospects of finding electroweak DM at future hadron colliders with the $14$-TeV and $100$-TeV centre of mass energies have been performed earlier with a detector design similar to that of the $8$ TeV Run-1 LHC~\cite{Low_Wang,Cirelli}. Possible interesting proposals for improving the reach of Higgsino-like DM in disappearing charged track search have also been put forward~\cite{Mahbubani,Hajime}. In particular, the latter studies investigated the impact of reducing the required number of hits in the tracking system by the candidate charged track, and found that under optimistic scenarios for the SM background estimates, the reach at a $33$~TeV~\cite{Hajime} and subsequently a $100$~TeV collider~\cite{Mahbubani} can be significantly improved for Higgsino states. 

In this paper, we examine the discovery potential of the electroweak DM at future high  energy hadron colliders. We consider two proposals: a 27-TeV upgrade of the LHC (HE-LHC), which can be achieved within the current LHC ring with upgraded magnets~\cite{FrankZ}, and the proposed 100 TeV future collider~\cite{Arkani-Hamed:2015vfh} at CERN (FCC-hh)~\cite{europe_100tev} and in China (SppC)~\cite{china_100tev}. 
We adopt the updated detector design with the new Insertable B-Layer (IBL) included in the ATLAS tracking system for the Run-2 $13$~TeV LHC~\cite{ATLAS_IBL}, and model our background estimates by extrapolating the ATLAS results, using similar methods as adopted in previous studies \cite{Low_Wang,Cirelli}. 

The subsequent sections are organized as follows. In Sec.~2 we briefly describe the relevant details of the electroweak DM model, the signal and background processes for the search channels utilized in our analysis, and the simulation framework adopted. In Sec.~3 we discuss the distribution of different kinematic observables used to distinguish between the signal and the background processes, the event selection criteria, and a simple optimization of the kinematic selections to improve the signal to background ratio. We then go on to present our main results, discussing the comparative reach of 14-TeV HL-LHC, 27-TeV HE-LHC and the 100-TeV FCC-hh/SppC options. We conclude with a brief summary of our results and an outlook in Sec.~4.

\section{Analysis Setup}

\subsection{Effective interaction Lagrangian}
\label{sec:Lag}
We begin with a brief review of the relevant effective interactions of pure wino and Higgsino states with the SM sector, as well as the radiative mass splitting between the charged and neutral components of the electroweakinos generated by SM gauge interactions. Although we will adopt the supersymmetric terminology to describe the ${\rm SU}(2)_{\rm L}$ doublet and triplet DM scenarios in the following, our discussion is valid in general for an effective theory, with the SM augmented by a stable DM multiplet. 

The effective interaction Lagrangian at dimension-4 for charged ($\tilde{\chi}^\pm$) and neutral ($\tilde{\chi}^0$) winos with the SM electroweak gauge bosons is given as
\begin{equation}
\mathcal{ L}^W_{V\chi\chi} \supseteq -g \left(\overline{\tilde{\chi}}^0 \gamma^\mu \tilde{\chi}^{+} W^{-}_\mu +{\rm h.c.} \right) + g \overline{\tilde{\chi}}^- \gamma^\mu \tilde{\chi}^{-} \left(\cos \theta_W Z_\mu +\sin \theta_W A_\mu \right),
\end{equation}
where $g$  is the ${\rm SU}(2)_{\rm L}$ gauge coupling, and $\theta_W$ is the weak-mixing angle. 
In the absence of large corrections from couplings with the fermion and sfermion sectors of the MSSM, these gauge interactions induce a mass splitting between the charged and neutral winos ($\delta m_{\tilde{W}}$), which, at the two-loop order can be parametrized as follows~\cite{Shigeki}
\begin{align}
\frac{\delta m_{\tilde{W}}}{1 {~\rm MeV}} & = -413.315 + 305.383 \left(\log \frac{m_{\tilde{\chi}_0}}{1 {~\rm GeV}}\right) - 60.8831 \left(\log \frac{m_{\tilde{\chi}_0}}{1 {~\rm GeV}}\right)^2  \\ \nonumber
 &  + 5.41948 \left(\log \frac{m_{\tilde{\chi}_0}}{1 {~\rm GeV}}\right)^3- 0.181509 \left(\log \frac{m_{\tilde{\chi}_0}}{1 {~\rm GeV}}\right)^4,
\end{align}
where $m_{\tilde{\chi}_0}$ is the neutral wino mass. The corresponding decay lifetime of the charged wino to a neutral wino and a charged pion is given in terms of the $c\tau$-value
by~\cite{Shigeki}
\begin{equation}
c\tau \simeq 3.1 {~\rm cm} \left[\left(\frac{\delta m_{\tilde{W}}}{164 {~\rm MeV}}\right)^{3} \sqrt{1- \frac{m_\pi^2}{\delta m_{\tilde{W}}^2}}\ \right]^{-1},
\label{eq:wino_split}
\end{equation}
with $m_\pi$ being the charged pion mass. We have normalized the mass difference to 164 MeV, which is the mass splitting in the limit of heavy WIMPs, $M_{\rm DM}\gg m_W$.

Similarly, the effective interaction Lagrangian at dimension-4 for charged (${\chi}_H^\pm$) and neutral (${\chi}_H^0$) Dirac Higgsinos with the SM electroweak gauge bosons is given by
\begin{align}
  {\cal L}^H_{V\chi\chi} & \supseteq  -\frac{g}{\sqrt{2}} \left( \overline{\chi^0_H}\gamma^\mu \chi^-_H\, W^+_\mu
+{\rm h.c.} \right)
+g \overline{\chi^-_H}\gamma^\mu \chi^-_H\,\left(\frac{1/2-s^2_W}{c_W}\,Z_\mu \, + s_W A_\mu \right) \nonumber \\
                                    & -  \frac{g}{2 c_W}\, \overline{\chi^0_H}\gamma^\mu \chi^0_H\, Z_\mu,
\end{align}
with $s_W=\sin\theta_W$ and $c_W=\cos\theta_W$.
The above interactions induce a one-loop mass splitting between the charged and neutral states ($\delta m_{\tilde{H}}$) which can be written as
\begin{equation}
\delta m_{\tilde{H}} = \frac{g^2}{16\pi^2}m_{\tilde{H}}\sin^2\theta_W f\left(\frac{m_Z}{m_{\tilde{H}}}\right),
\end{equation}
where the loop function is given by
\begin{equation}
f(r) = r^4\ln r - r^2 - r\sqrt{r^2 - 4}(r^2 + 2)\ln\frac{\sqrt{r^2 - 4} + r}{2}.
\end{equation}
The corresponding decay lifetime of the charged Higgsino to a neutral Higgsino and a charged pion can be parametrized in terms of the $c\tau$-value as~\cite{Hajime}
\begin{equation}
c\tau \simeq 0.7 {~\rm cm} \times \left [\left(\frac{\delta m_{\tilde{H}}}{340 {~\rm MeV}}  \right)^3 \sqrt{1-\frac{m_\pi^2}{\delta m^2_{\tilde{H}}}} \   \right]^{-1}.
\label{eq:higgsino_split}
\end{equation}
As we can observe from Eqs.~(\ref{eq:wino_split}) and (\ref{eq:higgsino_split}), for typical values of the mass splitting between the charged and neutral states, the charged wino has a considerably larger decay length compared to the charged Higgsino. This makes the search for winos more favorable than Higgsinos in the disappearing charged track analysis.

\subsection{Signal and background processes}
\label{sec:sig_bag}
As mentioned in the introduction, we will focus on two different search strategies for electroweak DM at hadron colliders, both of which are being carried out by the ATLAS and CMS collaborations at the Run-1 and Run-2 LHC: namely, the monojet plus missing transverse momentum search and the disappearing charged track analysis. For the 27-TeV HE-LHC and 100 TeV FCC-hh/SppC upgrades, we will also discuss a simple optimization of the kinematic selection criteria in the next section. In this section, we briefly describe the signal and background processes for these search channels, as well as the Monte Carlo (MC) simulation framework adopted for them in our analysis.

For the signal process, we consider the electroweak production of chargino and neutralino pair in proton-proton collisions, where the dominant contribution to the total cross-section comes from the following three sub-processes:
\begin{align}
 p p &\rightarrow \tilde{\chi}_1^+\tilde{\chi}_1^- +\text{jets},\nonumber \\
       &\rightarrow \tilde{\chi}_1^\pm\tilde{\chi}_i^0+\text{jets},\nonumber \\
       &\rightarrow \tilde{\chi}_i^0 \tilde{\chi}_j^0+\text{jets}, \text{~with~} i,j =1,2, \text{for~Higgsino-like~states}.
\end{align}        
Thus, in each event for the first two sub-processes either one or two charged states are produced, which is relevant for the disappearing track analysis.  All three sub-processes contribute to the signal in the monojet search channel, since the charged pions from the chargino decay are too soft to detect at hadron colliders~\footnote{Future electron-proton colliders, such as the LHeC or FCC-eh, could have unique sensitivity to BSM signals with such soft final state particles, and to short lifetimes of the decaying charged states~\cite{Curtin:2017bxr}.}. In the context of the MSSM, we have assumed here that all other sparticles except the DM multiplet is decoupled. Therefore, for the wino DM scenario, there is one light Majorana neutralino and one light chargino present in the low-energy spectrum. For the corresponding case of Higgsino DM, there is again one light chargino, while the number of light Majorana neutralinos is two. 

For both the search channels, with the presence of missing transverse momentum as one of the criteria, the dominant SM backgrounds come from single weak boson production in association with multiple hard jets. The total background cross-section in final states without any charged leptons is thus dominated by $Z+$jets production, with $Z\rightarrow\bar{\nu} \nu$. A similar order of magnitude contribution is obtained from $W^\pm+$jets production, with $W^+ \rightarrow \ell^+ \nu$ ($\ell=e,\mu,\tau$), where the charged lepton (except for hadronic decays of the tau) falls outside the tracker coverage of the detector. A sub-dominant component of the total background also comes from the top quark pair production process.

In our analysis, for both the signal and the background processes, we have generated MC events with matrix element (ME) and parton-shower matched using the MLM prescription~\cite{alpgen}, whereby we have included up to two additional hard jets at the ME level. All the parton level event samples have been obtained using the {\tt MG5aMCNLO}~\cite{MG5} event generator, followed by parton shower and hadronization with {\tt PYTHIA6.4.28}~\cite{Pythia}, and fast detector simulation using {\tt DELPHES3}~\cite{Delphes}. We have employed the {\tt CTEQ6L1}~\cite{cteq,LHAPDF} parton distribution functions, and have used the event-by-event default choice for the factorization and renormalization scales as implemented in {\tt MG5aMCNLO}. Jets have been defined using the {\tt anti}-$k_T$ algorithm~\cite{antikt,Fastjet}, with the radius parameter $R=0.5$. 

In order to obtain a large statistics of MC events in the kinematic region of our interest, we generated our event samples after strong cuts on the transverse momentum of the leading jets at the ME level. For the dominant as well as very large $Z+$jets background, we have applied an additional generation level cut on the missing transverse momentum variable. This makes it difficult for us to normalize our total matched cross-sections to next-to-leading order (NLO) in QCD results, since it requires a fully differential NLO simulation to obtain the proper K-factors after the above cuts. Therefore, we abstain from adopting a normalization by such K-factors for both the signal and background processes~\footnote{Since both the signal and background are electroweak processes, they have a similar NLO K-factor of about 1.4 \cite{Fox:2012ru, Kallweit:2015dum}. Therefore, we expect the higher order corrections not to change the signal to background ratio $S/B$ appreciably, but to slightly improve the statistical significance of the signal $S/\sqrt{B}$.}.

A few more comments are in order for the disappearing charged track analysis. Since our detector simulation cannot reproduce the trigger efficiency and the charged track selection efficiency obtained by the ATLAS collaboration, we have used an overall rescaling fudge factor of 0.1 to normalize our signal event yields to those reported by ATLAS \cite{wino_13tev_current}. 

Within our simulation framework, it is also difficult to estimate the SM background rates in the disappearing charged track analysis, which ensue from fake tracklets, missing leptons, and charged hadrons. We therefore adopt an empirical formula reported by the ATLAS collaboration by fitting the data obtained in their LHC Run-2 analysis. The differential distribution of the disappearing charged tracks as a function of their transverse momenta ($p_T$) can be parametrized as follows:
\begin{equation}
\frac{dN_{\rm Events}}{dp_T}=N_0 \exp  \left (-p_0  \log(p_T)  - p_1  \log(p_T)^2 \right),
\label{eq:fit}
\end{equation}
with $p_0=0.894$ and $p_1=0.057$, as obtained by a fit to the fake tracklet data from the 13 TeV LHC. The overall normalization factor $N_0$ is obtained to match the number of background events from fake tracklets obtained in the ATLAS analysis involving tracks with $p_T>20$~GeV. For different collider energies the functional dependence on the track $p_T$ is assumed to remain the same, while we rescale the overall normalization by the ratio of the $Z+$jets total cross-section, as also assumed in previous studies. Although $Z+$jets gives the dominant contribution to the total cross-section in the final state of interest (with a substantial missing transverse momentum and no charged lepton), there are also significant contributions from $W+$jets and $t\bar{t}+$jets processes. In order to take these latter contributions into account, along with the uncertainty in the background estimate using our simple methodology, we have varied the central value of the background rate by a factor of five in our subsequent analyses~\footnote{As we do not perform a shape analysis, only the total number of events after the relevant cut enters our final estimate. Although the actual shape is relevant in the cut optimization, we keep the shape fixed within the scope of our study.}.

\section{Results}

\begin{figure}[tb]
\centering
\includegraphics[scale=0.6]{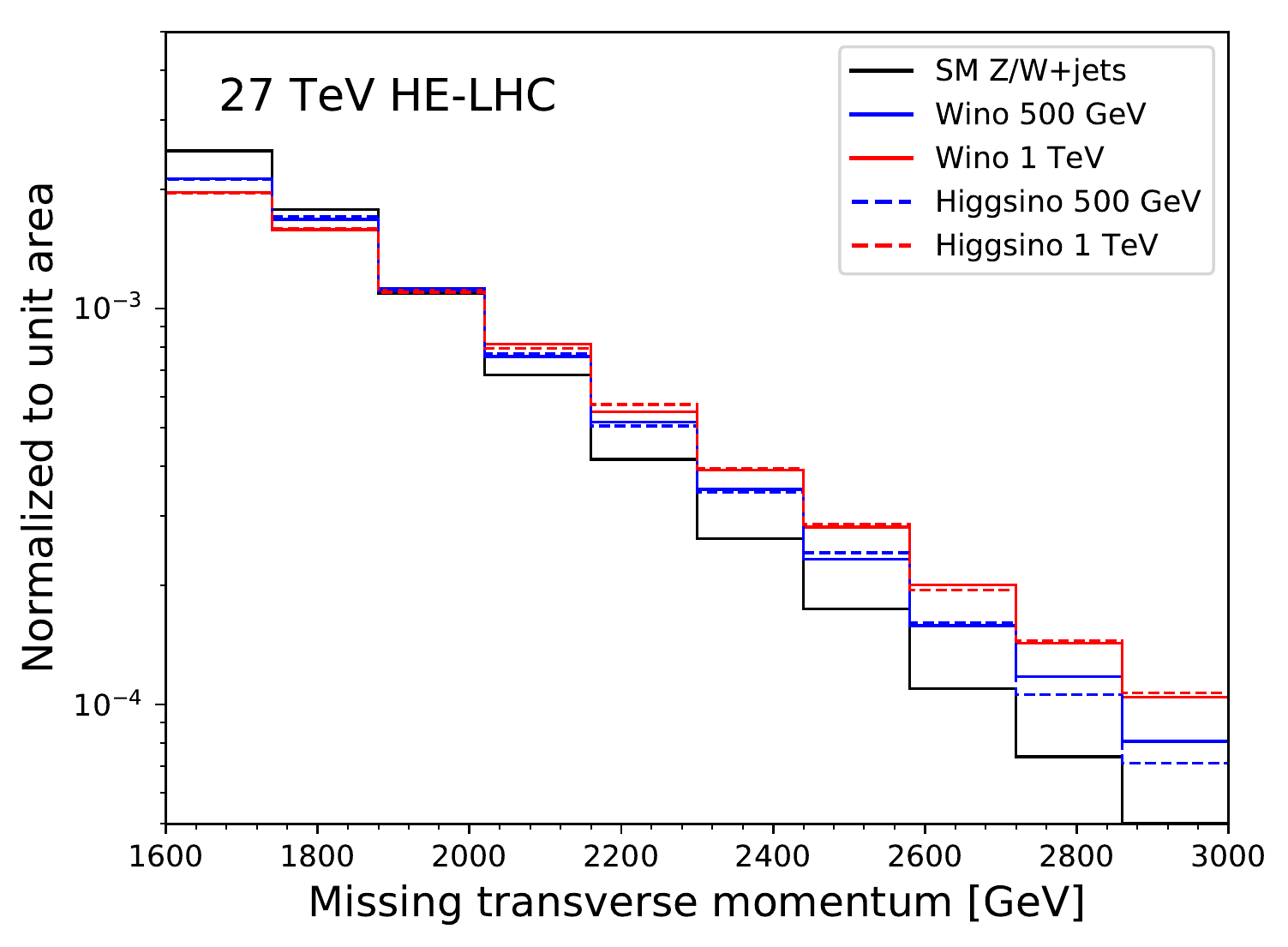}
\caption{\small Normalized distribution for missing transverse momentum in single weak boson ($W/Z$) production associated with multiple hard jets, in the SM (black solid lines), and in charged and neutral wino and Higgsino pair production events at the 27-TeV HE-LHC. The results are shown for two representative mass values of the winos (solid lines) and Higgsinos (dashed lines): 500 GeV (blue) and 1 TeV (red).}
\label{fig:monojet_met}
\end{figure}

\subsection{Kinematic selection of signal region}
\label{sec:selection}
\subsubsection{Mono-jet $+\slashed{E}_{T}$ search}
The kinematic selection criteria employed in the monojet plus missing transverse momentum channel is well-established, with increasingly stronger requirements on the transverse momenta of the hadronic jets (${p_T}_j$) and the missing transverse momentum ($\slashed{E}_{T}$), as the collider energy is increased. While we have optimized the above requirements for the HE-LHC and the FCC-hh/SppC analyses, for the corresponding HL-LHC scenario, our analysis closely follows the one by the CMS collaboration in Ref.~\cite{CMS:rwa}, to facilitate comparison. In our optimization of the kinematic selections for the higher centre of mass energies, we have maximized the statistical significance of the search, at the same time ensuring a large signal event rate.

In Fig.~\ref{fig:monojet_met}, we show the normalized distributions for missing transverse momentum at the $27$ TeV HE-LHC. The distributions are presented for both the signal process and the dominant SM background process of single weak boson ($W/Z$) production with multiple hard jets (black solid line). For the signal process, we show two representative mass values each for wino pair production (solid lines), and Higgsino pair production (dashed lines): namely, $500$~GeV (blue) and $1$~TeV (red). As we can see from this figure, for higher  mass of the winos and Higgsinos, the $p_T$ spectrum of the associated ISR (and hence the $\slashed{E}_{T}$) is also harder, as expected. A requirement of approximately $\slashed{E}_{T}>2$~TeV can potentially enhance the signal over background ratio for the above electroweakino mass values. 

For the event selection of the monojet channel, we require that all events have a hard central jet with a high threshold and also allow for a second jet with
\begin{equation}
p_T > p_{T,j_1}, \ \  |\eta| < 2; \ \ \ p_T > p_{T,j_2}, \ \ |\eta| < 4.5;\ \ 
{\rm and}\ \ \Delta R > 0.5, 
\end{equation}
and with an azimuthal separation $\Delta\phi_{j_1,j_2} < 2.5$, to remove back-to-back jets. 
Any additional jets passing the minimum threshold $p_T>p_{T,j_2}$ within $|\eta| < 4.5$ are vetoed, {\it i.e.,}
$N_{\rm jets} \leqslant 2$.
A lepton veto is applied with events with electrons (muons) with 
$p_T > 20~{\rm GeV}, \ \ |\eta| < 2.5~(2.1)$ are excluded. Events with hadronic taus with
$p_T > p_{T,\tau}, \ \ |\eta| < 2.4$ are also vetoed. Finally, an optimized requirement on missing transverse momentum is applied, with
$\slashed{E}_T>{\slashed{E}_T}^{\rm min}$.

We summarize the threshold values of the cuts, namely, ${\slashed{E}^{\rm min}_T},p_{T,j_1}, p_{T,j_2},  p_{T,\tau}$, for different collider options in Table~\ref{tab:monojets_cuts}. As mentioned earlier, we vary the $\slashed{E}_T$ and $p_{T,j_2}$ cuts for the 27~TeV and 100~TeV scenarios in the ranges specified in the table to optimize the signal significance.
\begin{table}
	\def\arraystretch{1.0}\centering
	\begin{tabular}{|c|c|c|c|c|}
		\hline
		$\sqrt{s}$ & $\slashed{E}^{\rm min}_T$ [GeV] & $p_{T,j_1}$~[GeV] & $p_{T,j_2}$~[GeV] & $p_{T,\tau}$~[GeV]\\
		\hline
		14 TeV & 650 & 300 & 30 & 30\\
		\hline
		27 TeV & 1800--2700 & 400 & 60--160 & 30\\
		\hline
		100 TeV & 4800--7000 & 1200 & 250--450 & 40\\
		\hline
	\end{tabular}
\caption{\small Threshold values of different kinematic observables, namely, ${\slashed{E}^{\rm min}_T},p_{T,j_1}, p_{T,j_2},  p_{T,\tau}$ for different collider options in the monojet analysis, and the optimization range considered for the HE-LHC and FCC-hh/SppC colliders. See text for details.}
\label{tab:monojets_cuts}

\end{table}

The optimized set of kinematic cuts for the HE-LHC is given in Table~\ref{Tab:monojet}, with the corresponding signal and background cross-sections. Here, basic cuts refers to the requirement of $\slashed{E}_T > 1600$~GeV at the matrix-element level. We also show the efficiency of each cut on the signal ($\epsilon_S$) and background rates ($\epsilon_B$), as well as the improvement in the signal-to-background ratio ($S/B$) with each cut. As we can see from Table~\ref{Tab:monojet}, for the representative mass value of $500$~GeV for the chargino and neutralino states, the $S/B$ ratio that can be achieved is at most $4.59 \times 10^{-2}$  for wino-like states, and $2.32 \times 10^{-2}$ for Higgsino-like states.

As such, the systematic errors could be a main concern for the monojet$+\slashed{E}_{T}$ search. However, encouragingly, the theoretical errors on the $W/Z+$jets background rates have been reduced to a few percent level with the recent NNLO QCD corrections and NLO electroweak corrections supplemented by Sudakov logarithms at two loops~\cite{WZ_theory}. At the same time, the current uncertainties on the estimate of the background cross-sections using data-driven methods are also at the few percent level~\cite{monojet_exp}, which are expected to further reduce with the accumulation of higher statistics. 

\begin{table}
\small
\def\arraystretch{1.0}\centering
\begin{tabular}{|c|c|c||c|c|c|c|c|c|}
\hline
\multirow{2}{*}{Cuts} &      Bckgrnd & \multirow{2}{*}{$\epsilon_{B}$} & \multicolumn{3}{|c|}{$m_{\tilde{W}} = 500$ GeV} & \multicolumn{3}{|c|}{$m_{\tilde{H}} = 500$ GeV} \\
\cline{4-6}
\cline{7-9}
                                         &            [fb] &                 &     Signal [fb] &  $\epsilon_{S}$ &      $S/B$~[\%] &     Signal [fb] &  $\epsilon_{S}$ &      $S/B$~[\%] \\
\hline
                              Basic cuts &           26.50 &               - &            0.40 &               - &            1.52 &            0.21 &               - &            0.79 \\
\hline
                  $p_{Tj_{1}} > 400$ GeV &           26.12 &            0.99 &            0.40 &            0.99 &            1.52 &            0.21 &            0.99 &            0.79 \\
\hline
              $N_{\rm jets} \leqslant 2$ &           21.13 &            0.81 &            0.33 &            0.83 &            1.55 &            0.17 &            0.83 &            0.81 \\
\hline
             $\Delta\phi_{j_{1}, j_{2}}$ &           20.13 &            0.95 &            0.32 &            0.98 &            1.60 &            0.17 &            0.98 &            0.83 \\
\hline
                               Muon veto &           16.13 &            0.80 &            0.32 &            1.00 &            1.99 &            0.17 &            1.00 &            1.04 \\
\hline
                           Electron veto &           12.78 &            0.79 &            0.32 &            1.00 &            2.52 &            0.17 &            1.00 &            1.31 \\
\hline
                                Tau veto &           10.88 &            0.85 &            0.31 &            0.98 &            2.88 &            0.16 &            0.98 &            1.50 \\
\hline
            $\slashed{E}_{T} > 2.2$ TeV &            1.03 &            0.09 &            0.05 &            0.15 &            4.59 &            0.02 &            0.15 &            2.32 \\
\hline
\end{tabular}
\caption{\small Signal and background cross-sections in the monojet$+\slashed{E}_{T}$ channel at the 
27-TeV HE-LHC after successive selection cuts on different kinematic observables; see text for details on the selection criteria. The efficiency of each cut on the signal ($\epsilon_S$) and background rates ($\epsilon_B$), along with the signal to background ratio ($S/B$) are also shown. We have shown the results for the representative mass value of $500$~GeV for the wino and Higgsino states.}

\label{Tab:monojet}
\end{table}

\subsubsection{Disappearing charged track search}
\begin{figure}[tb]
\centering
\includegraphics[scale=0.49]{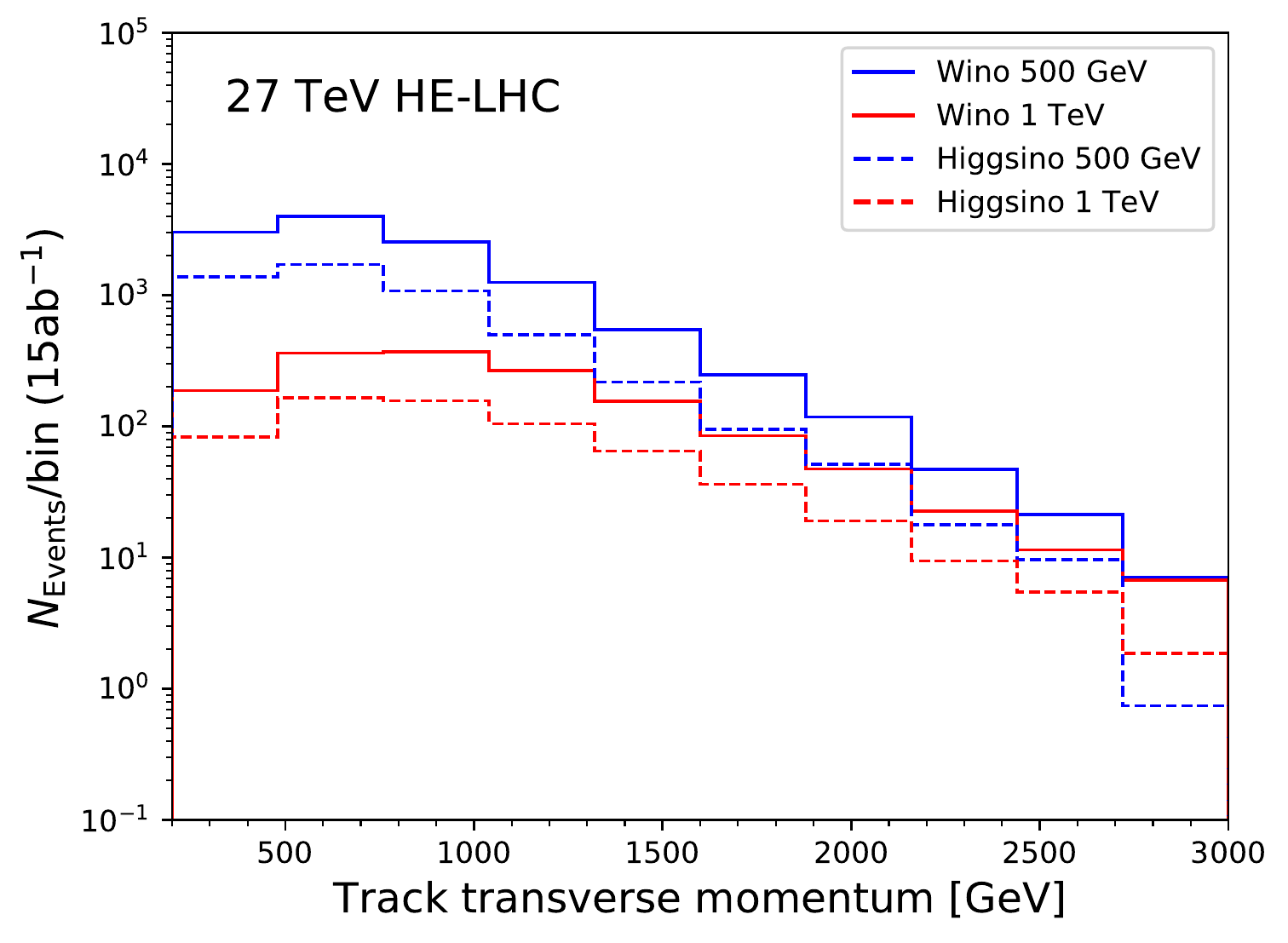}
\includegraphics[scale=0.49]{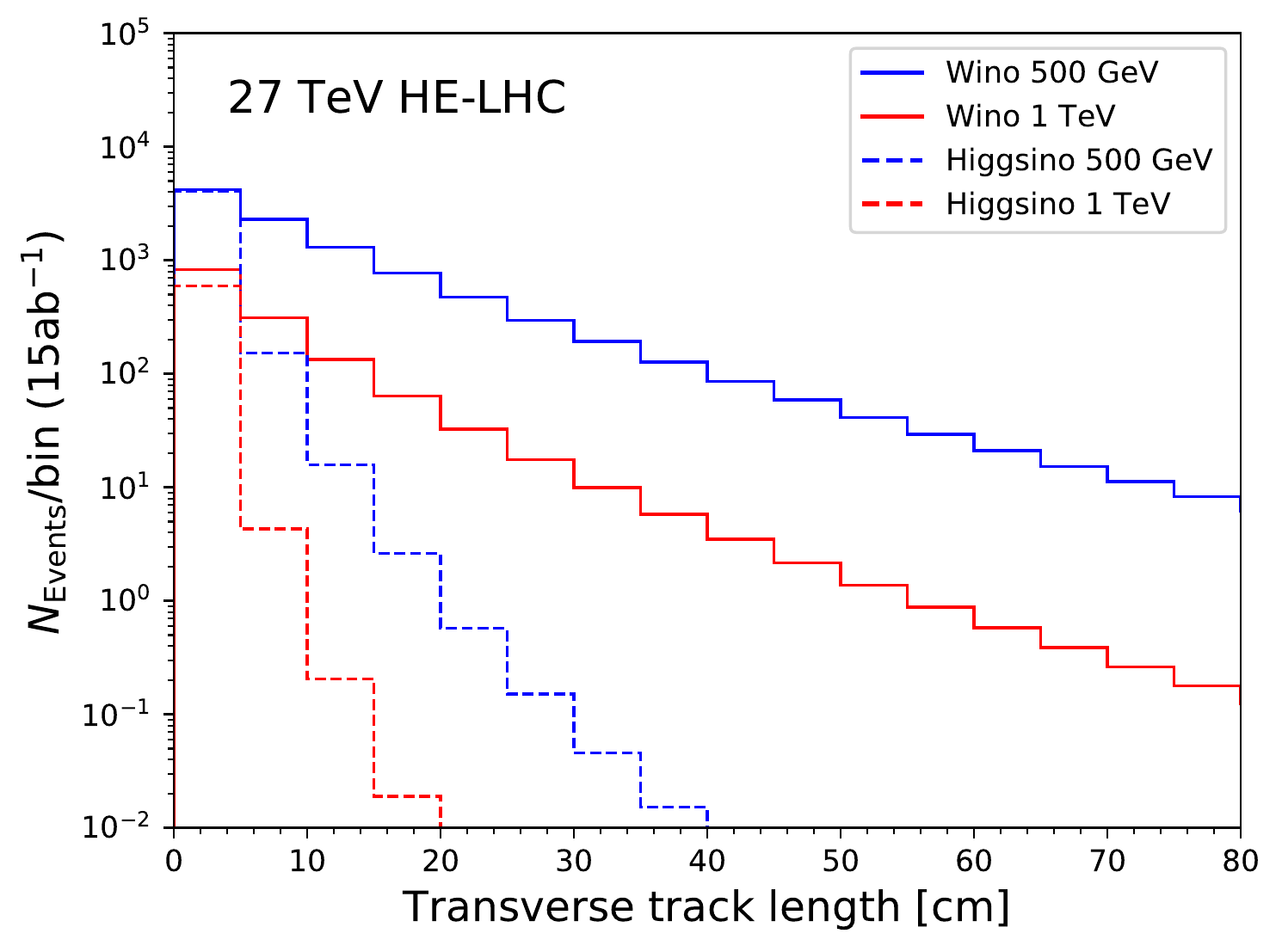}
\caption{\small  Transverse momentum (left panel) and transverse track length (right column) distribution of disappearing charged tracks in charged and neutral wino (solid lines) and Higgsino (dashed lines) pair production events (with at least one chargino in each event), for the 27-TeV HE-LHC with $15 {~\rm ab}^{-1}$ data. The results are shown for two representative mass values of the winos and Higgsinos: 500 GeV (blue) and 1 TeV (red).}
\label{fig:dist_track}
\end{figure}

We have discussed the methodology adopted for our estimate of the normalization and the shape of the SM backgrounds in the disappearing charged track search analysis in Sec.~\ref{sec:sig_bag}. We also described the lifetime (expressed as $c\tau$) of the charged wino and Higgsino states in their rest frame in Sec.~\ref{sec:Lag}. The decay length of the charginos in the LHC detectors is 
determined by $c\tau$ and the transverse momentum distribution of the chargino. We show in Fig.~\ref{fig:dist_track} (left column) the transverse momentum distribution of the chargino track for both the wino-like (solid line) and the Higgsino-like (dashed line) scenarios at the $27$ TeV HE-LHC with $15 {~\rm ab}^{-1}$ data. The distributions have been shown for the chargino mass value of $500$~GeV (blue) and $1$~TeV (red). As we can see from this figure, the overall shape of the distribution is similar for Higgsino and wino-like states, while the total production cross-section is a factor of two larger in the latter scenario. 

Combined with the proper life-time, the transverse momentum distribution of the tracks determine the transverse charged track length in the signal events, which is the most important observable in the disappearing charged track analysis. We show this distribution in Fig.~\ref{fig:dist_track} (right column), with the parameter choice and color coding used same as for the previous figure. It is clear from this figure that in order to probe a Higgsino of mass $\mathcal{O}(1 {~\rm TeV})$, we need tracking coverage in the range of $10-20$ cm, which is now possible after the inclusion of the additional B-layer in the Run-2  upgrade of the ATALS detector. 

We now briefly describe the event selection criteria used for the disappearing charged track analysis. We require one hard central jet  plus large missing momentum in the events with
\begin{equation}
p_T > p_{T,j_1}, \ \  |\eta| < 2.8,\quad \slashed{E}_T>{\slashed{E}_T}^{\rm min}.
\end{equation}
Charged lepton veto is applied as described above for the monojet channel. Furthermore, the missing transverse momentum vector is required to have an azimuthal separation from the leading jet by
\begin{equation}
\Delta\phi_{j_1, \slashed{\vec{E}}_{T}} > 1.5.
\end{equation}
If there is a second jet with
\begin{equation}
p_T > p_{T,j_2}, \ \  |\eta| < 2.8,
\end{equation}
in addition, $\Delta\phi_{j_2, \slashed{\vec{E}}_{T}} > 1.5$ is also required. A candidate charged track is required to have
\begin{equation}
p_T > p_{T \rm track}, \ \  0.1<|\eta|<0.9,
\end{equation}
with no hadronic jet within a cone of $\Delta R < 0.4$, where $\Delta R$ is the separation in the pseudorapidity azimuthal angle plane. Finally, we demand all events to have at least one candidate track with radial track length in the range
\begin{equation}
12 < d < 30~{\rm cm}.
\end{equation}
We summarize the threshold values of the cuts, namely, ${\slashed{E}^{\rm min}_T},p_{T,j_1}, p_{T,j_2}$ and  $p_{T,\rm track}$, for different collider options in Table~\ref{Tab:track_cuts}. As mentioned earlier, we vary the $\slashed{E}_T, p_{T,j_1}$ and $p_{T,\rm track}$ cuts for the 27~TeV and 100~TeV scenarios in the ranges specified in the table to optimize the signal significance.

\begin{table}
	\def\arraystretch{1.0}\centering
	\begin{tabular}{|c|c|c|c|c|}
		\hline
		$\sqrt{s}$ & $\slashed{E}_T$ [GeV] & $p_{T,j_1}$~[GeV] & $p_{T,j_2}$~[GeV] & $p_{T,\rm track}$~[GeV]\\
		\hline
		14 TeV & 150 & 150 & 70 & 250\\
		\hline
		27 TeV & 400 -- 700 & 400 -- 600 & 140 & 400 -- 700\\
		\hline
		100 TeV & 1000 -- 1400 & 700 -- 1400 & 500 & 1000 -- 1400\\
		\hline
	\end{tabular}
\caption{\small Threshold values of different kinematic observables, namely, ${\slashed{E}^{\rm min}_T},p_{T,j_1}, p_{T,j_2}$ and  $p_{T,\rm track}$ for different collider options in the disappearing charged track analysis, and the optimization range considered for the HE-LHC and FCC-hh/SppC colliders. See text for details.}
\label{Tab:track_cuts}
\end{table}

The optimized set of kinematic cuts for the HE-LHC is given in Table~\ref{Tab:track_opti}, with the corresponding signal cross-sections. Here, basic cuts refers to the requirement of $\slashed{E}_T > 150$~GeV at the matrix-element level. We also show the efficiency of each cut on the signal ($\epsilon_S$)  rates.  As we can see from Table~\ref{Tab:monojet}, for the representative mass value of $500$~GeV for the chargino and neutralino states in the wino-like scenario, we expect a cross-section of $1.59$~fb, which, after taking into account the efficiency fudge factor of $0.1$ mentioned in Sec.~\ref{sec:sig_bag}, would imply $2385$~signal events with $15 {~\rm ab}^{-1}$ data at the HE-LHC. Following the methodology described in the above section, we also expect around $28$ background events. Thus, even if the background normalization increases by upto a factor of five, the signal to background ratio, $S/B$, would be in the range of $17-85$. Similarly, for the Higgsino-like scenario, the $S/B$ ratio is estimated to be in the range of $1-7$ for the representative mass value of $300$~GeV. Both these numbers are encouraging and imply that with a detector design similar to that of Run-2 LHC, the experimental uncertainties in the disappearing charged track search will be largely statistical in nature. 

\begin{table}
\def\arraystretch{1.0}\centering
\begin{tabular}{|c|c|c|c|c|}
\hline
\multirow{2}{*}{Cuts} & \multicolumn{2}{|c|}{$m_{\tilde{W}} = 500$ GeV} & \multicolumn{2}{|c|}{$m_{\tilde{H}} = 300$ GeV} \\
\cline{2-3}
\cline{4-5}
                                         &     Signal [fb] &  $\epsilon_{S}$ &     Signal [fb] &  $\epsilon_{S}$ \\
\hline
                              Basic cuts &          102.91 &               - &          242.53 &               - \\
\hline
                             Lepton veto &          102.90 &            1.00 &          242.52 &            1.00 \\
\hline
                  $p_{Tj_{1}} > 450$ GeV &           16.42 &            0.16 &           30.86 &            0.13 \\
\hline
             $\slashed{E}_{T} > 550$ GeV &           11.29 &            0.69 &           18.81 &            0.61 \\
\hline
 $\Delta\phi_{j, \slashed{E}_{T}} > 1.5$ &           10.61 &            0.94 &           17.54 &            0.93 \\
\hline
             $p_{T \rm track} > 400$ GeV & \multirow{2}{*}{7.05} & \multirow{2}{*}{0.66} & \multirow{2}{*}{8.43} & \multirow{2}{*}{0.48} \\
                         Track isolation &                 &                 &                 &                 \\
\hline
                       $ 12 < d < 30$ cm &            1.59 &            0.23 &            0.13 &            0.01 \\
\hline
\end{tabular}
\caption{\small Signal cross-section in the disappearing charged track analysis at the 27-TeV HE-LHC after successive selection  cuts on the kinematic and track-quality observables; see text for details on the selection criteria. The efficiency of each cut on the signal ($\epsilon_S$)  rates is also shown. We have presented the results for the representative mass value of $500$~GeV ($300$~GeV) for the wino (Higgsino) states.}
\label{Tab:track_opti}
\end{table}

\subsection{Comparative reach of different hadron collider options}
\label{sec:compare}
We are now in a position to compare the reach of different hadron collider options in searching for wino and Higgsino dark matter and their associated charged states. We will show the results for three different scenarios of the collider energy and integrated luminosity:
\begin{eqnarray}
&{\rm HL\text{-}LHC:} \quad &14\tev,\quad 3~{\rm ab}^{-1}, \nonumber \\
&{\rm HE\text{-}LHC:} \quad &27\tev,\quad 15~{\rm ab}^{-1}, \nonumber \\
&{\rm FCC\text{-}hh/SppC:} \quad &100\tev,\quad 30~{\rm ab}^{-1}.
\end{eqnarray}

\begin{figure}[tb]
\centering
\includegraphics[scale=0.461]{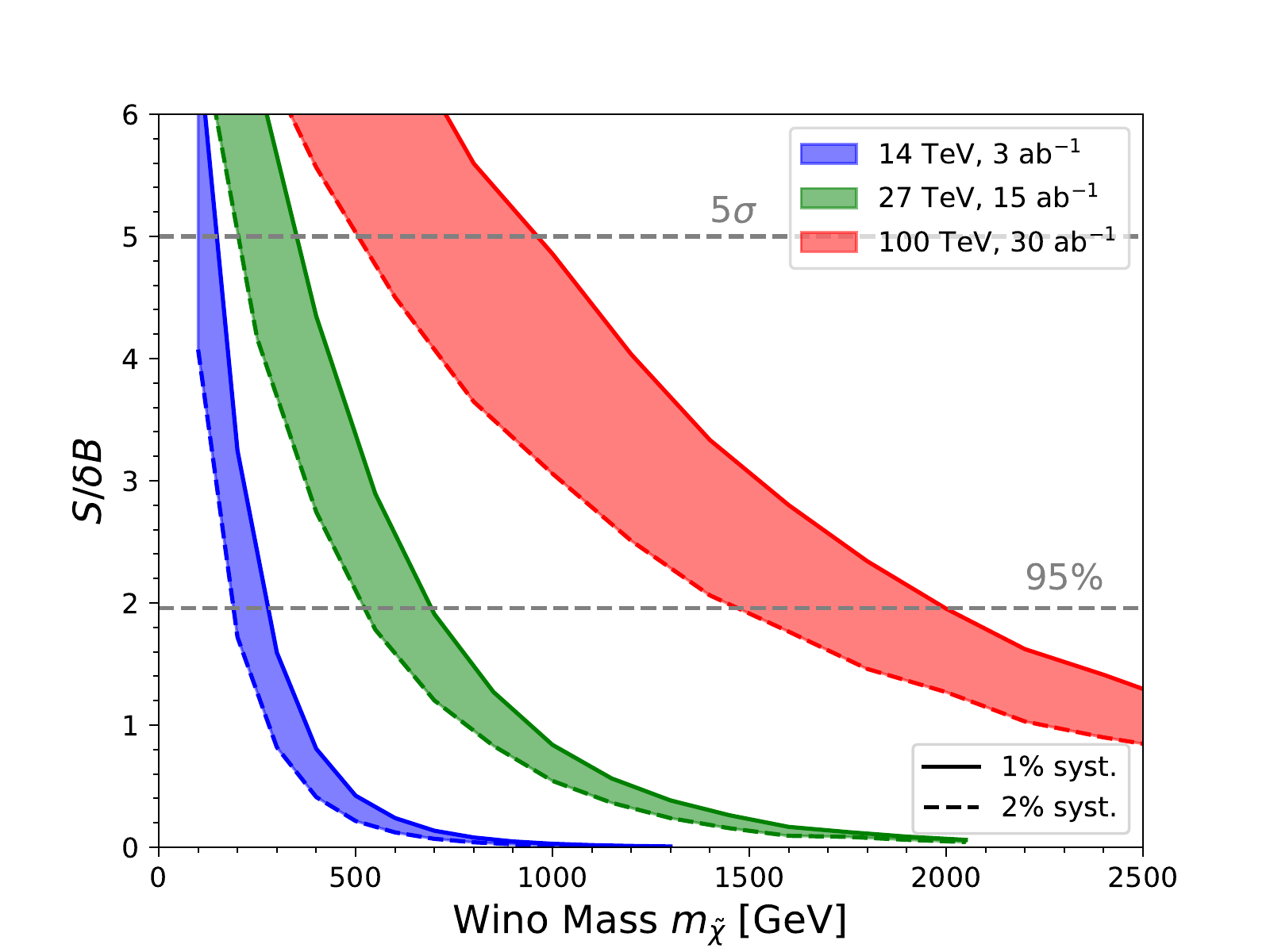}
\includegraphics[scale=0.461]{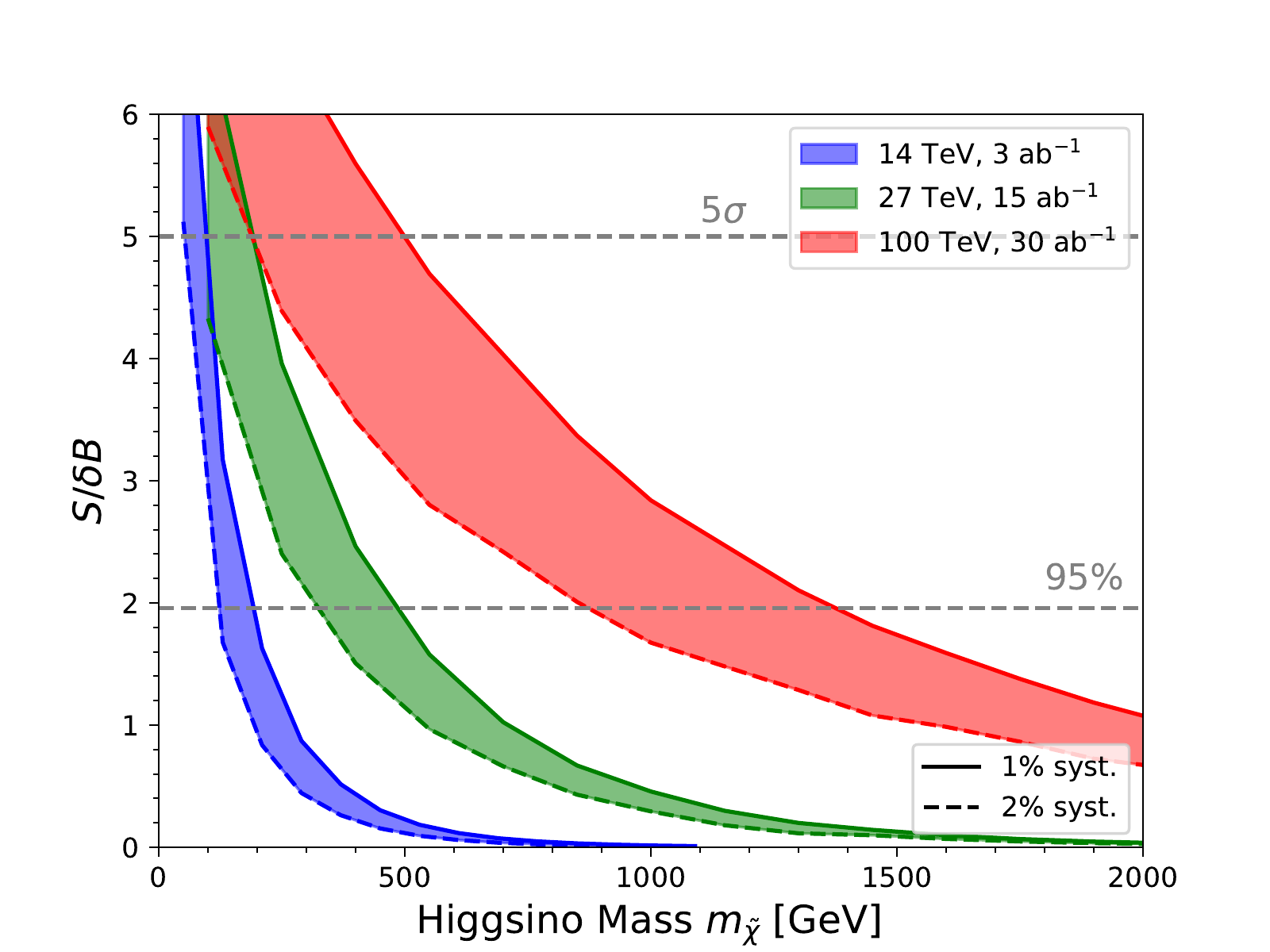}
\caption{\small Comparative reach of the HL-LHC, HE-LHC and FCC-hh/SppC options in the monojet channel for wino-like (left panel) and Higgsino-like (right panel) DM search. The solid and dashed lines correspond to optimistic values of the systematic uncertainties on the background estimate of $1\%$ and $2\%$ respectively, which might be achievable using data-driven methods with the accumulation of large statistics.}
\label{fig:reach_monojet}
\end{figure}

To present our results on the future reach of the above collider options, we adopt a definition of significance 
\begin{equation}
S\over {\sqrt{B+(\Delta_B B)^2 + (\Delta_S S)^2 }}, 
\end{equation}
where $S$ and $B$ are the total number of signal and background events as before, and 
$\Delta_S, \Delta_B$ refer to the corresponding percentage systematic uncertainties, respectively. For the monojet channel, we have taken $\Delta_B=1-2\%$ and $\Delta_S=10\%$, while for the disappearing charged track analysis, we assume $\Delta_B=20\%$ and $\Delta_S=10\%$. As emphasized earlier, although the systematic uncertainties in the current LHC analyses in the above channels are larger, the uncertainties in the background estimate using data-driven methods are expected to further reduce with the accumulation of higher statistics. Furthermore, since our background estimate in the disappearing track analysis is a simple extrapolation of the ATLAS results for the $13$~TeV LHC, we have also varied the central value of the background yield within a factor of five (i.e., between $20\%$ and $500\%$) of the number obtained using the method discussed in Sec.~\ref{sec:sig_bag}.

In Fig.~\ref{fig:reach_monojet} we compare the reach of the HL-LHC, HE-LHC and FCC-hh/SppC options in the monojet channel for wino-like (left panel) and Higgsino-like (right panel) DM search, where $\delta B = \sqrt{B+(\Delta_B B)^2 + (\Delta_S S)^2 }$. The solid and dashed lines correspond to systematic uncertainties on the background estimate of $1\%$ and $2\%$ respectively.  In an optimistic scenario, we can expect to probe at the $95\%$ C.L. wino-like DM mass of upto $280, 700$ and $2000$~GeV, at the $14,27$ and $100$~TeV colliders respectively. For the Higgsino-like scenario, these numbers are reduced to $200,490$ and $1370$~GeV, primarily due to the reduced production cross-section. Clearly, a $27$~TeV collider can achieve a substantially improved reach by a factor of two or more compared to the HL-LHC, while the $100$~TeV collider option will improve it further by another factor of three. Furthermore, a $100$~TeV collider option may be able to completely cover the thermal Higgsino mass window using the monojet search, if the systematic uncertainties can be brought down to a percent level.

\begin{figure}[tb]
\centering
\includegraphics[scale=0.46]{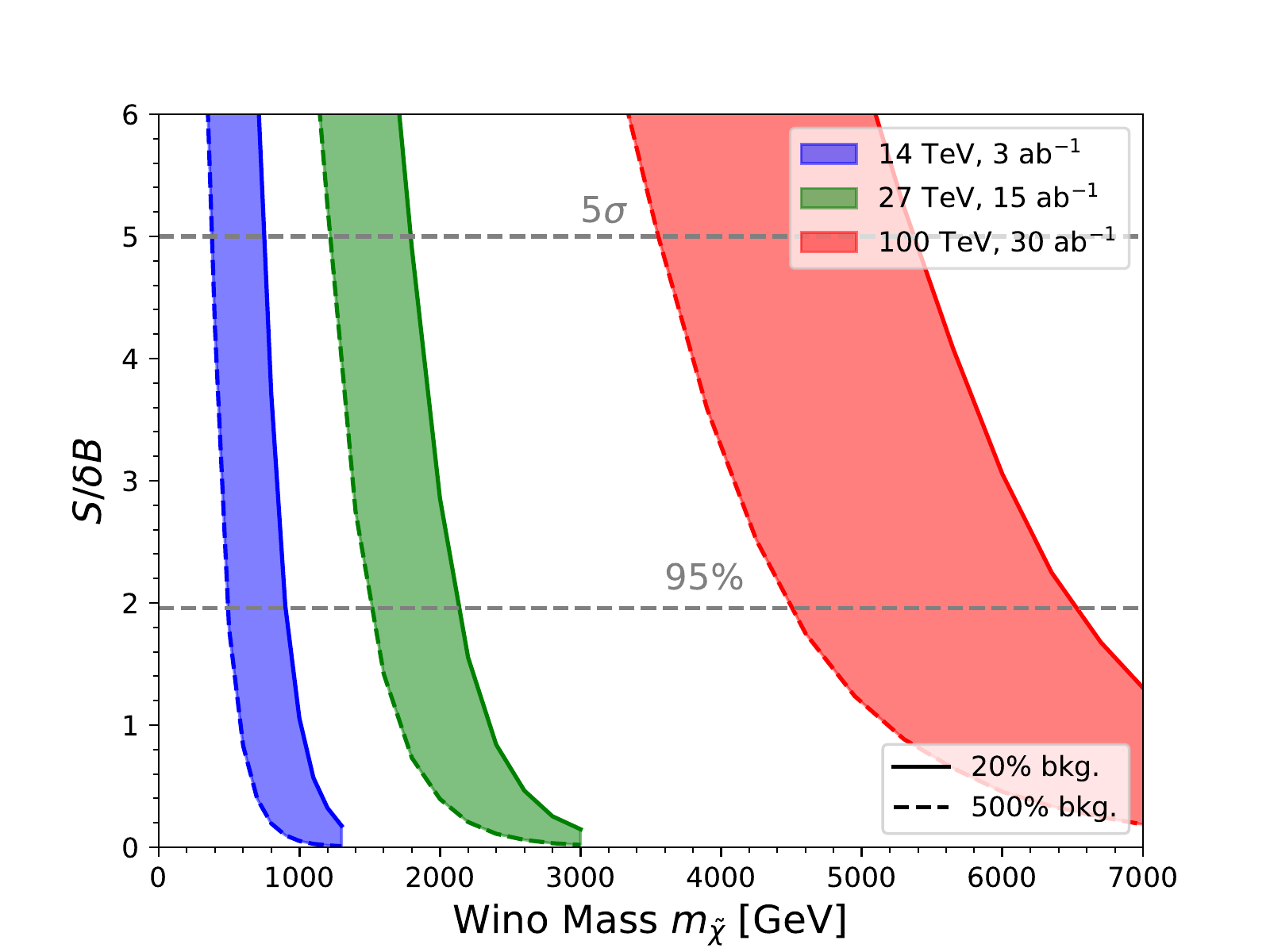}
\includegraphics[scale=0.46]{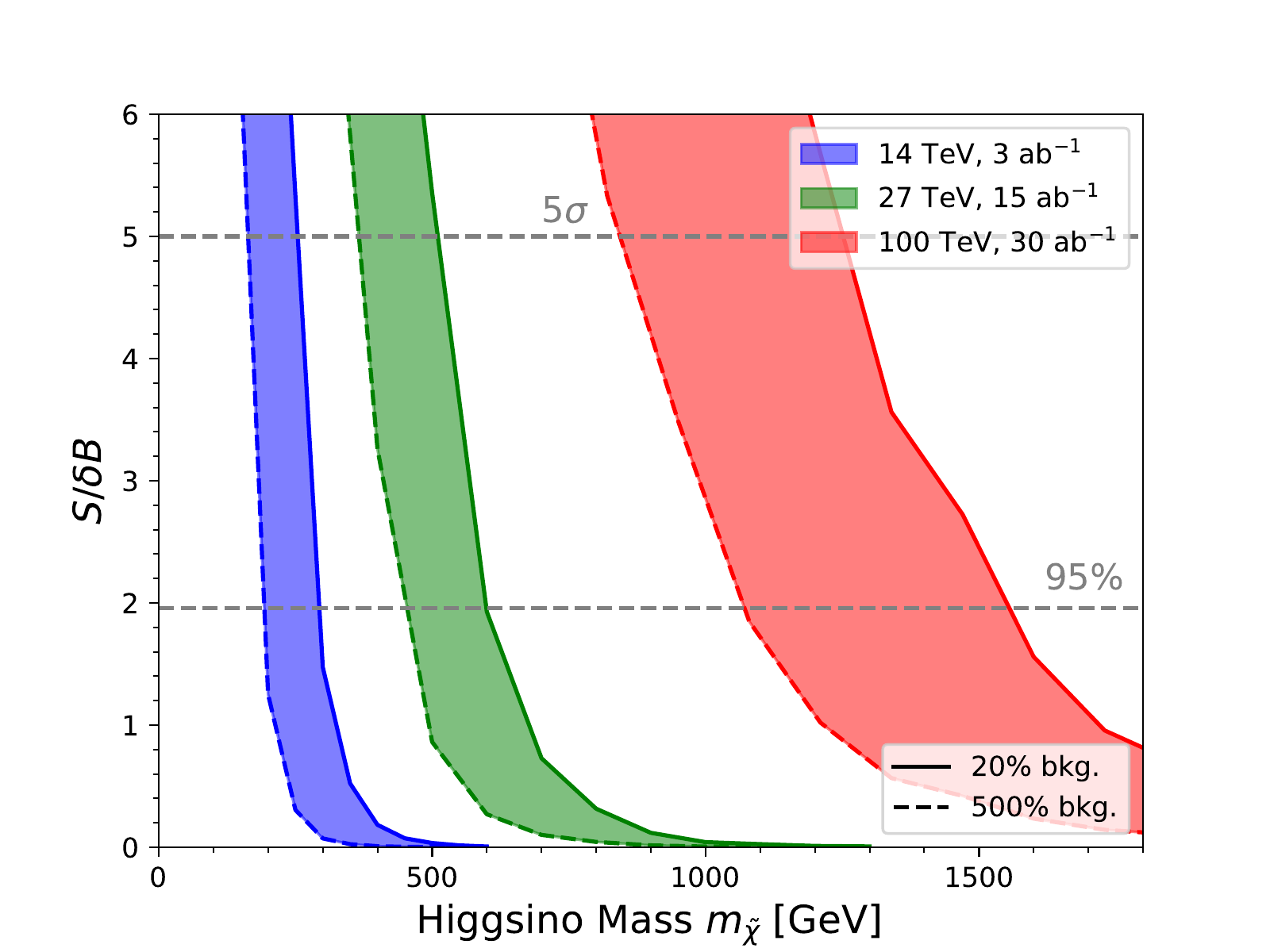}
\caption{\small Comparative reach of the HL-LHC, HE-LHC and FCC-hh/SppC options in the disappearing charged track analysis for wino-like (left panel) and Higgsino-like (right panel) DM search. The solid and dashed lines correspond to modifying the central value of the background estimate by a factor of five, i.e., $20\%$ and $500\%$ of that obtained through the fit function in Eq.~\ref{eq:fit}.}
\label{fig:reach_track}
\end{figure}

In Fig.~\ref{fig:reach_track} we compare the reach of the HL-LHC, HE-LHC and FCC-hh/SppC options in the disappearing charged track analysis for wino-like (left panel) and Higgsino-like (right panel) DM search. The solid and dashed lines correspond to modifying the central value of the background estimate by a factor of five, i.e., $20\%$ and $500\%$ of that obtained through the fit function in Eq.~\eqref{eq:fit}. With the lower value of the background estimate, the expected reach on wino-like DM mass at the $95\%$ C.L. is $0.9,2.1$ and $6.5$~TeV at the $14,27$ and $100$~TeV colliders respectively. For the Higgsino-like scenario, these numbers are reduced to $300,600$ and $1550$~GeV, primarily due to the smaller length of the disappearing track and the reduced production rate. For the higher value of the background estimate, the mass reach for the wino-like states are modified to $500,1500$ and $4500$~GeV, respectively, at the three collider energies. Similarly, for the Higgsino-like scenario, the reach is modified to $200,450$ and $1070$~GeV. We note that the signal significance in the disappearing track search is rather sensitive to the wino and Higgsino mass values (thus making the $2\sigma$ and $5\sigma$ reach very close in mass). This is because, as the chargino lifetime in the lab frame becomes shorter for heavier masses, the signal event rate decreases exponentially.

The improvements in going from the HL-LHC to the HE-LHC, and further from the HE-LHC to the FCC-hh/SppC are very similar to those obtained for the monojet analysis above, namely, around a factor of two and three, respectively. Although we have presented the reach at the $100$~TeV collider without reference to the cosmology of these DM candidates, in order for a wino heavier than around $3$~TeV and a Higgsino heavier than around $1$~TeV not to overclose the Universe, one would require a non-standard thermal history, with late-time entropy production~\cite{Hooper}.

\section{Summary and Outlook}
Among the multitude of possibilities for particle DM, WIMPs remain a highly motivated candidate due to the predictable nature of the thermal relic abundance, and the correlated predictions for their experimental and observational probes. WIMP dark matter particles that belong to a multiplet of the standard model weak interactions are one of the best representatives, but are often challenging to probe in direct detection experiments due to loop-suppressed scattering cross-sections. Searches at hadron colliders are thus crucial for testing such a scenario, and depending upon the gauge representation, can be complementary to indirect detection probes in different mass windows. Moreover, since the relic abundance of electroweak DM is uniquely determined by its mass value, they represent a well-defined target in the  collider search for DM in general. 

In this paper, we studied collider probes of two representative scenarios for electroweak DM, namely an wino-like SU(2)$_{\rm L}$ triplet and a Higgsino-like SU(2)$_{\rm L}$ doublet. In the absence of higher-dimensional operators, radiative corrections generate a small mass splitting between the charged and neutral components of these multiplets, of the order of a few hundred MeV, as reviewed in Sec.~\ref{sec:Lag}. This nearly degenerate spectrum motivates two major search channels at hadron colliders for electroweak DM and its charged counterparts, namely, the monojet with missing transverse momentum search and the disappearing charged track analysis. We examined the relevant signal and background processes for these search channels in proton-proton collisions in Sec.~\ref{sec:sig_bag}, along with the methodology adopted for estimating the event rates at colliders.  

\begin{table}
	\def\arraystretch{1.0}\centering
	\begin{tabular}{|c|c|c|c|c|}
		\hline
		95\% & Wino  & Wino & Higgsino & Higgsino \\		
		C.L. &  Monojet & Disappearing Track &  Monojet & Disappearing  Track \\
		\hline
		14 TeV &  280 GeV& 900 GeV & 200 GeV  &   300 GeV\\
		\hline
		27 TeV & 700 GeV & 2.1 TeV & 490 GeV & 600 GeV\\
		\hline
		100 TeV &  2 TeV&  6.5 TeV & 1.4 TeV &  1.5 TeV \\
		\hline
	\end{tabular}
\caption{\small Summary of DM mass reach at 95$\%$ C.L. for an electroweak triplet (wino-like) and a doublet (Higgsino-like) representation, at the HL-LHC, HE-LHC and the FCC-hh/SppC colliders, in optimistic scenarios for the background systematics. See text for details.}
\label{tab:summary}
\end{table}

We presented the distributions of the important kinematic observables and the details of the kinematic selection criteria followed at different collider energies in Sec.~\ref{sec:selection}. For our analysis, we considered three options for future hadron colliders: the high-luminosity HL-LHC, the proposed 27-TeV LHC upgrade (HE-LHC) and the 100-TeV FCC-hh/SppC. We performed an optimization of the selection criteria for the two higher centre of mass energies, maximizing the statistical significance of the particular search, at the same time ensuring a large signal event rate.

The estimates of the expected mass reach at the 27-TeV HE-LHC are discussed in Sec.~\ref{sec:compare}. We also presented comparisons with the projected reach for the 14-TeV HL-LHC and the 100-TeV hadron collider. Our results for these three options are summarized in Table~\ref{tab:summary}. In particular, we find that the disappearing charged track analysis at the HE-LHC can probe Higgsino-like (wino-like) DM mass of up to $600$~GeV ($2.1$~TeV) at the $95\%$ C.L., making it complementary to the indirect probes using gamma rays from dwarf-spheroidal galaxies, as mentioned in Sec.~\ref{sec:intro}. The monojet and missing transverse momentum search, on the otherhand, has a weaker reach of $490$~GeV ($700$~GeV) at $95\%$ C.L. for the Higgsino-like (wino-like) states. We further see in Table~\ref{tab:summary} that across different collider energies, while the reach for wino-like states is improved in the disappearing track analysis by around a factor of three compared to the monojet search, for Higgsino-like states the improvement is of the order of 100~GeV. We note that the performance in the monojet search will crucially depend on how far the systematic uncertainties can be reduced using data-driven methods at future high-luminosity runs, as the signal to background ratio remains at the few percent level.

For the disappearing charged track analysis, we adopted a detector setup similar to that of the ATLAS tracking system for the Run-2 LHC upgrade, with a new Insertable B-Layer (IBL), which crucially extends the search reach for Higgsino-like states with a shorter decay length in the tracker. Since the only way to understand the backgrounds for this search is from the data, we extrapolated the 13 TeV ATLAS results to higher energies, keeping the shape of the distribution as a function of the track transverse momentum unchanged, while normalizing by the ratio of the total rates at different energies. 

Although quite representative, it should be noted that our studies are limited to the case of pure electroweak doublet and triplet states. In more general scenarios, in which mixing among the electroweak multiplets, in particular, with an additional fermionic gauge singlet is non-negligible, the considerations of relic abundance and detection techniques would be substantially altered. This can also lead to rather rich physics scenarios at colliders, depending upon the other particles in the spectrum and their mass differences. For example, with a non-degenerate spectra in the chargino-neutralino sector of the MSSM, decays of the heavier states can produce electroweak gauge bosons, which would in turn lead to signals with multiple charged leptons and missing transverse momenta in the final state. 

The LHC and its high-luminosity upgrade will lead the research in the energy frontier for the coming decades. The possible high-energy upgrade to 27 TeV, the HE-LHC, is an exciting option, and a potentially important step towards the 100-TeV territory at the FCC-hh/SppC colliders. As we found in this paper, the proposed HE-LHC could significantly extend the scope of electroweak DM searches beyond the reach of the HL-LHC. It would thus have a fantastic potential for discovery, providing boost to the future collider programme at 100 TeV.

%%%%%%%%%%%%%%%%%%%%%%%%%%
\begin{center} \textbf{Acknowledgment} \end{center}

We would like to thank Matthew Low and Lian-Tao Wang for discussions. This work was supported in part by the U.S.~Department of Energy under grant No.~DE-FG02- 95ER40896 and by the PITT PACC. TH also acknowledges the hospitality of the Aspen Center for Physics, which is supported by National Science Foundation grant PHY-1607611.

%%%%%%%%%%%%%%%%%%%%%%%%%%%%%%%%%%%%%%%%%%%%%%%%%%%%%%%%%%%%


\begin{thebibliography}{99}
\bibitem{strumia_minimal}
%\cite{Cirelli:2005uq}
%\bibitem{Cirelli:2005uq} 
  M.~Cirelli, N.~Fornengo and A.~Strumia,
  %``Minimal dark matter,''
  Nucl.\ Phys.\ B {\bf 753}, 178 (2006).
%  doi:10.1016/j.nuclphysb.2006.07.012
 % [hep-ph/0512090].
  %%CITATION = doi:10.1016/j.nuclphysb.2006.07.012;%%
  
\bibitem{strumia_minimal2}
%\cite{Cirelli:2007xd}
%\bibitem{Cirelli:2007xd} 
  M.~Cirelli, A.~Strumia and M.~Tamburini,
  %``Cosmology and Astrophysics of Minimal Dark Matter,''
  Nucl.\ Phys.\ B {\bf 787}, 152 (2007).
%  doi:10.1016/j.nuclphysb.2007.07.023
%  [arXiv:0706.4071 [hep-ph]].
  %%CITATION = doi:10.1016/j.nuclphysb.2007.07.023;%%
  %384 citations counted in INSPIRE as of 29 Apr 2018  
  
\bibitem{susy_dm_review}
%\cite{Jungman:1995df}
%\bibitem{Jungman:1995df} 
  G.~Jungman, M.~Kamionkowski and K.~Griest,
  %``Supersymmetric dark matter,''
  Phys.\ Rept.\  {\bf 267}, 195 (1996).
%  doi:10.1016/0370-1573(95)00058-5
 % [hep-ph/9506380];
  %%CITATION = doi:10.1016/0370-1573(95)00058-5;%%  
  
  

%\cite{Lee:1977ua}
\bibitem{Lee:1977ua} 
  B.~W.~Lee and S.~Weinberg,
  %``Cosmological Lower Bound on Heavy Neutrino Masses,''
  Phys.\ Rev.\ Lett.\  {\bf 39}, 165 (1977).
%  doi:10.1103/PhysRevLett.39.165
  %%CITATION = doi:10.1103/PhysRevLett.39.165;%%
  %1080 citations counted in INSPIRE as of 21 Apr 2018
  
  %\cite{Goldberg:1983nd}
\bibitem{Goldberg:1983nd} 
  H.~Goldberg,
  %``Constraint on the Photino Mass from Cosmology,''
  Phys.\ Rev.\ Lett.\  {\bf 50}, 1419 (1983)
  Erratum: [Phys.\ Rev.\ Lett.\  {\bf 103}, 099905 (2009)].
%  doi:10.1103/PhysRevLett.103.099905, 10.1103/PhysRevLett.50.1419
  %%CITATION = doi:10.1103/PhysRevLett.103.099905, 10.1103/PhysRevLett.50.1419;%%
  %1256 citations counted in INSPIRE as of 21 Apr 2018
  
 %\cite{Steigman:2012nb}
\bibitem{Steigman:2012nb} 
  G.~Steigman, B.~Dasgupta and J.~F.~Beacom,
  %``Precise Relic WIMP Abundance and its Impact on Searches for Dark Matter Annihilation,''
  Phys.\ Rev.\ D {\bf 86}, 023506 (2012).
%  doi:10.1103/PhysRevD.86.023506
 % [arXiv:1204.3622 [hep-ph]].
  %%CITATION = doi:10.1103/PhysRevD.86.023506;%%
  %307 citations counted in INSPIRE as of 21 Apr 2018 
  
\bibitem{Griest}
%\cite{Griest:1990kh}
%\bibitem{Griest:1990kh} 
  K.~Griest and D.~Seckel,
  %``Three exceptions in the calculation of relic abundances,''
  Phys.\ Rev.\ D {\bf 43}, 3191 (1991).
%  doi:10.1103/PhysRevD.43.3191
  %%CITATION = doi:10.1103/PhysRevD.43.3191;%%
  
\bibitem{Gondolo}
%\cite{Gondolo:1990dk}
%\bibitem{Gondolo:1990dk} 
  P.~Gondolo and G.~Gelmini,
  %``Cosmic abundances of stable particles: Improved analysis,''
  Nucl.\ Phys.\ B {\bf 360}, 145 (1991).
%  doi:10.1016/0550-3213(91)90438-4
  %%CITATION = doi:10.1016/0550-3213(91)90438-4;%%
  
\bibitem{Edsjo}
%\cite{Edsjo:1997bg}
%\bibitem{Edsjo:1997bg} 
  J.~Edsjo and P.~Gondolo,
  %``Neutralino relic density including coannihilations,''
  Phys.\ Rev.\ D {\bf 56}, 1879 (1997).
%  doi:10.1103/PhysRevD.56.1879
%  [hep-ph/9704361].
  %%CITATION = doi:10.1103/PhysRevD.56.1879;%%
  
\bibitem{well_tempered}
%\cite{ArkaniHamed:2006mb}
%\bibitem{ArkaniHamed:2006mb} 
  N.~Arkani-Hamed, A.~Delgado and G.~F.~Giudice,
  %``The Well-tempered neutralino,''
  Nucl.\ Phys.\ B {\bf 741}, 108 (2006).
 % doi:10.1016/j.nuclphysb.2006.02.010
 % [hep-ph/0601041].
  %%CITATION = doi:10.1016/j.nuclphysb.2006.02.010;%%
  
\bibitem{shigeki_sommerfeld}
%\cite{Hisano:2003ec}
%\bibitem{Hisano:2003ec} 
  J.~Hisano, S.~Matsumoto and M.~M.~Nojiri,
  %``Explosive dark matter annihilation,''
  Phys.\ Rev.\ Lett.\  {\bf 92}, 031303 (2004);
%  doi:10.1103/PhysRevLett.92.031303
%  [hep-ph/0307216].
  %%CITATION = doi:10.1103/PhysRevLett.92.031303;%%
%\cite{Hisano:2004ds}
%\bibitem{Hisano:2004ds} 
  J.~Hisano, S.~Matsumoto, M.~M.~Nojiri and O.~Saito,
  %``Non-perturbative effect on dark matter annihilation and gamma ray signature from galactic center,''
  Phys.\ Rev.\ D {\bf 71}, 063528 (2005);
 % doi:10.1103/PhysRevD.71.063528
 % [hep-ph/0412403].
  %%CITATION = doi:10.1103/PhysRevD.71.063528;%%
%\cite{Hisano:2006nn}
%\bibitem{Hisano:2006nn} 
  J.~Hisano, S.~Matsumoto, M.~Nagai, O.~Saito and M.~Senami,
  %``Non-perturbative effect on thermal relic abundance of dark matter,''
  Phys.\ Lett.\ B {\bf 646}, 34 (2007).
%  doi:10.1016/j.physletb.2007.01.012
 % [hep-ph/0610249].
  %%CITATION = doi:10.1016/j.physletb.2007.01.012;%%

  
  
\bibitem{Higgsino_coannihilation}
%\cite{Mizuta:1992qp}
%\bibitem{Mizuta:1992qp} 
  S.~Mizuta and M.~Yamaguchi,
  %``Coannihilation effects and relic abundance of Higgsino dominant LSP(s),''
  Phys.\ Lett.\ B {\bf 298}, 120 (1993).
%  doi:10.1016/0370-2693(93)91717-2
 % [hep-ph/9208251].
  %%CITATION = doi:10.1016/0370-2693(93)91717-2;%%
  
\bibitem{Moroi} 
%\cite{Moroi:1999zb}
%\bibitem{Moroi:1999zb} 
T.~Moroi and L.~Randall,
%``Wino cold dark matter from anomaly mediated SUSY breaking,''
Nucl.\ Phys.\ B {\bf 570}, 455 (2000).
%doi:10.1016/S0550-3213(99)00748-8
%[hep-ph/9906527].
%%CITATION = doi:10.1016/S0550-3213(99)00748-8;%%
%515 citations counted in INSPIRE as of 20 Apr 2018
  


\bibitem{Hisano_DD_Wino}
%\cite{Hisano:2010fy}
%\bibitem{Hisano:2010fy} 
  J.~Hisano, K.~Ishiwata and N.~Nagata,
  %``A complete calculation for direct detection of Wino dark matter,''
  Phys.\ Lett.\ B {\bf 690}, 311 (2010);
%  doi:10.1016/j.physletb.2010.05.047
 % [arXiv:1004.4090 [hep-ph]].
  %%CITATION = doi:10.1016/j.physletb.2010.05.047;%%
%\cite{Hisano:2011cs}
%\bibitem{Hisano:2011cs} 
  J.~Hisano, K.~Ishiwata, N.~Nagata and T.~Takesako,
  %``Direct Detection of Electroweak-Interacting Dark Matter,''
  JHEP {\bf 1107}, 005 (2011);
%  doi:10.1007/JHEP07(2011)005
%  [arXiv:1104.0228 [hep-ph]].
  %%CITATION = doi:10.1007/JHEP07(2011)005;%%
%\cite{Hisano:2012wm}
%\bibitem{Hisano:2012wm} 
  J.~Hisano, K.~Ishiwata and N.~Nagata,
  %``Direct Search of Dark Matter in High-Scale Supersymmetry,''
  Phys.\ Rev.\ D {\bf 87}, 035020 (2013);
 % doi:10.1103/PhysRevD.87.035020
%  [arXiv:1210.5985 [hep-ph]].
  %%CITATION = doi:10.1103/PhysRevD.87.035020;%%
  %\cite{Hisano:2015rsa}
%\bibitem{Hisano:2015rsa} 
  J.~Hisano, K.~Ishiwata and N.~Nagata,
  %``QCD Effects on Direct Detection of Wino Dark Matter,''
  JHEP {\bf 1506}, 097 (2015).
%  doi:10.1007/JHEP06(2015)097
%  [arXiv:1504.00915 [hep-ph]].
  %%CITATION = doi:10.1007/JHEP06(2015)097;%%


\bibitem{DD_Hill}
%\cite{Hill:2011be}
%\bibitem{Hill:2011be} 
  R.~J.~Hill and M.~P.~Solon,
  %``Universal behavior in the scattering of heavy, weakly interacting dark matter on nuclear targets,''
  Phys.\ Lett.\ B {\bf 707}, 539 (2012);
%  doi:10.1016/j.physletb.2012.01.013
%  [arXiv:1111.0016 [hep-ph]].
  %%CITATION = doi:10.1016/j.physletb.2012.01.013;%%
%\cite{Hill:2013hoa}
%\bibitem{Hill:2013hoa} 
  R.~J.~Hill and M.~P.~Solon,
  %``WIMP-nucleon scattering with heavy WIMP effective theory,''
  Phys.\ Rev.\ Lett.\  {\bf 112}, 211602 (2014).
%  doi:10.1103/PhysRevLett.112.211602
%  [arXiv:1309.4092 [hep-ph]].
  %%CITATION = doi:10.1103/PhysRevLett.112.211602;%%


%\cite{TuckerSmith:2001hy}
\bibitem{dm_inelastic} 
%\cite{Han:1997wn}
%\bibitem{Han:1997wn} 
  T.~Han and R.~Hempfling,
  %``Messenger sneutrinos as cold dark matter,''
  Phys.\ Lett.\ B {\bf 415}, 161 (1997);
%  doi:10.1016/S0370-2693(97)01205-7
%  [hep-ph/9708264].
  %%CITATION = doi:10.1016/S0370-2693(97)01205-7;%%
D.~Tucker-Smith and N.~Weiner,
%``Inelastic dark matter,''
Phys.\ Rev.\ D {\bf 64}, 043502 (2001).
%doi:10.1103/PhysRevD.64.043502
%[hep-ph/0101138].
%%CITATION = doi:10.1103/PhysRevD.64.043502;%%
%598 citations counted in INSPIRE as of 20 Apr 2018


\bibitem{wino_indirect_shigeki}
%\cite{Bhattacherjee:2014dya}
%\bibitem{Bhattacherjee:2014dya} 
  B.~Bhattacherjee, M.~Ibe, K.~Ichikawa, S.~Matsumoto and K.~Nishiyama,
  %``Wino Dark Matter and Future dSph Observations,''
  JHEP {\bf 1407}, 080 (2014).
%  doi:10.1007/JHEP07(2014)080
%  [arXiv:1405.4914 [hep-ph]].
  %%CITATION = doi:10.1007/JHEP07(2014)080;%%
  
 \bibitem{Shigeki_review}
 %\cite{Matsumoto:2017znc}
%\bibitem{Matsumoto:2017znc} 
  S.~Matsumoto,
  %``Unexplored regions of WIMP,''
  PoS KMI {\bf 2017}, 033 (2017).
%  doi:10.22323/1.294.0033
  %%CITATION = doi:10.22323/1.294.0033;%%
 
 \bibitem{Reece}
 %\cite{Krall:2017xij}
%\bibitem{Krall:2017xij} 
  R.~Krall and M.~Reece,
  %``Last Electroweak WIMP Standing: Pseudo-Dirac Higgsino Status and Compact Stars as Future Probes,''
  Chin.\ Phys.\ C {\bf 42}, no. 4, 043105 (2018).
%  doi:10.1088/1674-1137/42/4/043105
%  [arXiv:1705.04843 [hep-ph]].
  %%CITATION = doi:10.1088/1674-1137/42/4/043105;%%
 


  
\bibitem{ATLAS_track_8TeV}
%\cite{TheATLAScollaboration:2013bia}
%\bibitem{TheATLAScollaboration:2013bia} 
  The ATLAS collaboration [ATLAS Collaboration],
  %``Search for charginos nearly mass-degenerate with the lightest neutralino based on a disappearing-track signature in $pp$ collisions at $\sqrt{s}=8~\mathrm{TeV}$ with the ATLAS detector,''
  ATLAS-CONF-2013-069.
  %%CITATION = ATLAS-CONF-2013-069;%%

  
  
\bibitem{wino_13tev_current}  
%\cite{Aaboud:2017mpt}
%\bibitem{Aaboud:2017mpt} 
  M.~Aaboud {\it et al.} [ATLAS Collaboration],
  %``Search for long-lived charginos based on a disappearing-track signature in $pp$ collisions at $\sqrt{s}$ = 13 TeV with the ATLAS detector,''
  arXiv:1712.02118 [hep-ex].
  %%CITATION = ARXIV:1712.02118;%%  

\bibitem{higgsino_13tev_current}  
  The ATLAS collaboration [ATLAS Collaboration],
%``Search for direct pair production of higgsinos by reinterpretation of the disappearing track analysis with 36.1~$fb^{−1}$ of $\sqrt{s} = 13$~TeV data collected with the ATLAS experiment,''
ATL-PHYS-PUB-2017-019. 


\bibitem{CMS_8TeV}
%\cite{CMS:2014gxa}
%\bibitem{CMS:2014gxa} 
  V.~Khachatryan {\it et al.} [CMS Collaboration],
  %``Search for disappearing tracks in proton-proton collisions at $ \sqrt{s}=8 $ TeV,''
  JHEP {\bf 1501}, 096 (2015).
%  doi:10.1007/JHEP01(2015)096
%  [arXiv:1411.6006 [hep-ex]].
  %%CITATION = doi:10.1007/JHEP01(2015)096;%%
  
\bibitem{CMS_13TeV}
%\cite{Sirunyan:2018ldc}
%\bibitem{Sirunyan:2018ldc} 
  A.~M.~Sirunyan {\it et al.} [CMS Collaboration],
  %``Search for disappearing tracks as a signature of new long-lived particles in proton-proton collisions at $\sqrt{s} =$ 13 TeV,''
  arXiv:1804.07321 [hep-ex].
  %%CITATION = ARXIV:1804.07321;%%
  



\bibitem{Low_Wang}
%\cite{Low:2014cba}
%\bibitem{Low:2014cba} 
  M.~Low and L.~T.~Wang,
  %``Neutralino dark matter at 14 TeV and 100 TeV,''
  JHEP {\bf 1408}, 161 (2014).
%  doi:10.1007/JHEP08(2014)161
%  [arXiv:1404.0682 [hep-ph]].
  %%CITATION = doi:10.1007/JHEP08(2014)161;%%
  
 \bibitem{Cirelli}
 %\cite{Cirelli:2014dsa}
%\bibitem{Cirelli:2014dsa} 
  M.~Cirelli, F.~Sala and M.~Taoso,
  %``Wino-like Minimal Dark Matter and future colliders,''
  JHEP {\bf 1410}, 033 (2014)
  Erratum: [JHEP {\bf 1501}, 041 (2015)].
%  doi:10.1007/JHEP10(2014)033, 10.1007/JHEP01(2015)041
%  [arXiv:1407.7058 [hep-ph]].
  %%CITATION = doi:10.1007/JHEP10(2014)033, 10.1007/JHEP01(2015)041;%%
 
 
\bibitem{Mahbubani}  
%\cite{Mahbubani:2017gjh}
%\bibitem{Mahbubani:2017gjh} 
  R.~Mahbubani, P.~Schwaller and J.~Zurita,
  %``Closing the window for compressed Dark Sectors with disappearing charged tracks,''
  JHEP {\bf 1706}, 119 (2017)
  Erratum: [JHEP {\bf 1710}, 061 (2017)].
%  doi:10.1007/JHEP06(2017)119, 10.1007/JHEP10(2017)061
%  [arXiv:1703.05327 [hep-ph]].
  %%CITATION = doi:10.1007/JHEP06(2017)119, 10.1007/JHEP10(2017)061;%%

\bibitem{Hajime}
%\cite{Fukuda:2017jmk}
%\bibitem{Fukuda:2017jmk} 
  H.~Fukuda, N.~Nagata, H.~Otono and S.~Shirai,
  %``Higgsino Dark Matter or Not: Role of Disappearing Track Searches at the LHC and Future Colliders,''
  Phys.\ Lett.\ B {\bf 781}, 306 (2018).
%  doi:10.1016/j.physletb.2018.03.088
%  [arXiv:1703.09675 [hep-ph]].
  %%CITATION = doi:10.1016/j.physletb.2018.03.088;%%
  
  
\bibitem{FrankZ}
Talk presented by Michaell Benedikt and Frank Zimmermann at the {\tt HL-LHC/HE-LHC Physics Workshop}, CERN, Oct.~2017. 

  %\cite{Arkani-Hamed:2015vfh}
\bibitem{Arkani-Hamed:2015vfh} 
  N.~Arkani-Hamed, T.~Han, M.~Mangano and L.~T.~Wang,
  %``Physics opportunities of a 100 TeV proton�proton collider,''
  Phys.\ Rept.\  {\bf 652}, 1 (2016).
%  doi:10.1016/j.physrep.2016.07.004
 % [arXiv:1511.06495 [hep-ph]].
  %%CITATION = doi:10.1016/j.physrep.2016.07.004;%%
  %120 citations counted in INSPIRE as of 11 Feb 2018
  %
 
  \bibitem{europe_100tev}  
  T.~Golling {\it et al.},
%``Physics at a 100 TeV pp collider: beyond the Standard Model phenomena,''
  CERN Yellow Report, no. 3, 441 (2017)  [arXiv:1606.00947 [hep-ph]];
  %%CITATION = doi:10.23731/CYRM-2017-003.441;%%
  %64 citations counted in INSPIRE as of 02 Feb 2018
  R.~Contino {\it et al.},
  %``Physics at a 100 TeV pp collider: Higgs and EW symmetry breaking studies,''
  CERN Yellow Report, no. 3, 255 (2017)
  %doi:10.23731/CYRM-2017-003.255
  [arXiv:1606.09408 [hep-ph]].
  %%CITATION = doi:10.23731/CYRM-2017-003.255;%%
  %73 citations counted in INSPIRE as of 02 Feb 2018

\bibitem{china_100tev} 
CEPC-SPPC Preliminary Conceptual Design Report, The CEPC-SPPC Study Group, IHEP- CEPC-DR-2015-01 (2015).

  
\bibitem{ATLAS_IBL}
%\cite{Capeans:2010jnh}
%\bibitem{Capeans:2010jnh} 
  M.~Capeans {\it et al.} [ATLAS Collaboration],
  %``ATLAS Insertable B-Layer Technical Design Report,''
  CERN-LHCC-2010-013, ATLAS-TDR-19.
  %%CITATION = CERN-LHCC-2010-013, ATLAS-TDR-19;%%
  


\bibitem{Shigeki}
%\cite{Ibe:2012sx}
%\bibitem{Ibe:2012sx} 
  M.~Ibe, S.~Matsumoto and R.~Sato,
  %``Mass Splitting between Charged and Neutral Winos at Two-Loop Level,''
  Phys.\ Lett.\ B {\bf 721}, 252 (2013).
%  doi:10.1016/j.physletb.2013.03.015
%  [arXiv:1212.5989 [hep-ph]].
  %%CITATION = doi:10.1016/j.physletb.2013.03.015;%%
  %84 citations counted in INSPIRE as of 16 Feb 2018
    
    
%\cite{Curtin:2017bxr}
\bibitem{Curtin:2017bxr} 
D.~Curtin, K.~Deshpande, O.~Fischer and J.~Zurita,
%``New Physics Opportunities for Long-Lived Particles at Electron-Proton Colliders,''
arXiv:1712.07135 [hep-ph].
%%CITATION = ARXIV:1712.07135;%%
%6 citations counted in INSPIRE as of 14 May 2018
  
  
\bibitem{alpgen}  
%\cite{Mangano:2002ea}
%\bibitem{Mangano:2002ea} 
  M.~L.~Mangano, M.~Moretti, F.~Piccinini, R.~Pittau and A.~D.~Polosa,
  %``ALPGEN, a generator for hard multiparton processes in hadronic collisions,''
  JHEP {\bf 0307}, 001 (2003).
%  doi:10.1088/1126-6708/2003/07/001
%  [hep-ph/0206293].
  %%CITATION = doi:10.1088/1126-6708/2003/07/001;%%

\bibitem{MG5}
%\cite{Alwall:2011uj}
%\bibitem{Alwall:2011uj} 
  J.~Alwall, M.~Herquet, F.~Maltoni, O.~Mattelaer and T.~Stelzer,
%  ``MadGraph 5 : Going Beyond,''
  JHEP {\bf 1106}, 128 (2011);
%  [arXiv:1106.0522 [hep-ph]].
  %%CITATION = ARXIV:1106.0522;%%
  %703 citations counted in INSPIRE as of 15 Aug 2013
  %\cite{Alwall:2014hca}
%\bibitem{Alwall:2014hca}
  J.~Alwall {\it et al.},
 % ``The automated computation of tree-level and next-to-leading order differential cross sections, and their matching to parton shower simulations,''
  JHEP {\bf 1407} (2014) 079.
%  doi:10.1007/JHEP07(2014)079
 % [arXiv:1405.0301 [hep-ph]].
  %%CITATION = doi:10.1007/JHEP07(2014)079;%%

%
\bibitem{Pythia}
%\cite{Sjostrand:2006za}
%\bibitem{Sjostrand:2006za} 
  T.~Sjostrand, S.~Mrenna and P.~Z.~Skands,
 % ``PYTHIA 6.4 Physics and Manual,''
  JHEP {\bf 0605}, 026 (2006).
%  [hep-ph/0603175].
  %%CITATION = HEP-PH/0603175;%%
  %4295 citations counted in INSPIRE as of 15 Aug 2013
  
    \bibitem{Delphes}
%\cite{Ovyn:2009tx}
%\bibitem{Ovyn:2009tx}
  S.~Ovyn, X.~Rouby and V.~Lemaitre,
%  ``DELPHES, a framework for fast simulation of a generic collider experiment,''
  arXiv:0903.2225 [hep-ph];
  %%CITATION = ARXIV:0903.2225;%%
  %\cite{deFavereau:2013fsa}
%\bibitem{deFavereau:2013fsa}
  J.~de Favereau {\it et al.} [DELPHES 3 Collaboration],
%  ``DELPHES 3, A modular framework for fast simulation of a generic collider experiment,''
  JHEP {\bf 1402} (2014) 057.
  %doi:10.1007/JHEP02(2014)057
  %[arXiv:1307.6346 [hep-ex]].
  %%CITATION = doi:10.1007/JHEP02(2014)057;%%
  
  

%\cite{Pumplin:2002vw}
\bibitem{cteq}
  J.~Pumplin, D.~R.~Stump, J.~Huston, H.~L.~Lai, P.~M.~Nadolsky and W.~K.~Tung,
 % ``New generation of parton distributions with uncertainties from global QCD analysis,''
  JHEP {\bf 0207} (2002) 012.
%  doi:10.1088/1126-6708/2002/07/012
 % [hep-ph/0201195].
  %%CITATION = doi:10.1088/1126-6708/2002/07/012;%%

  
    

%
\bibitem{LHAPDF}
%\cite{Whalley:2005nh}
%\bibitem{Whalley:2005nh}
  M.~R.~Whalley, D.~Bourilkov and R.~C.~Group,
 % ``The Les Houches accord PDFs (LHAPDF) and LHAGLUE,''
  hep-ph/0508110.
  %%CITATION = HEP-PH/0508110;%%


 \bibitem{antikt}
%\cite{Cacciari:2008gp}
%\bibitem{Cacciari:2008gp}
  M.~Cacciari, G.~P.~Salam and G.~Soyez,
  %``The Anti-k(t) jet clustering algorithm,''
  JHEP {\bf 0804} (2008) 063.
%  [arXiv:0802.1189 [hep-ph]].
  %%CITATION = ARXIV:0802.1189;%%
% 
%


\bibitem{Fastjet}
%\cite{Cacciari:2011ma}
%\bibitem{Cacciari:2011ma
  M.~Cacciari, G.~P.~Salam and G.~Soyez,
 % ``FastJet user manual,''
  Eur.\ Phys.\ J.\ C {\bf 72} (2012) 1896;
%  [arXiv:1111.6097 [hep-ph]].
  %%CITATION = ARXIV:1111.6097;%%
%\cite{hep-ph/0512210}
%\bibitem{hep-ph/0512210}
  M.~Cacciari and G.~P.~Salam,
%  ``Dispelling the $N^{3}$ myth for the $k_t$ jet-finder,''
  Phys.\ Lett.\ B\ {\bf 641} (2006) 57.
%  [hep-ph/0512210].
  %%CITATION = PHLTA,B641,57;%%


%\cite{Fox:2012ru}
\bibitem{Fox:2012ru} 
  P.~J.~Fox and C.~Williams,
  %``Next-to-Leading Order Predictions for Dark Matter Production at Hadron Colliders,''
  Phys.\ Rev.\ D {\bf 87}, no. 5, 054030 (2013)
  doi:10.1103/PhysRevD.87.054030
  [arXiv:1211.6390 [hep-ph]].
  %%CITATION = doi:10.1103/PhysRevD.87.054030;%%
  %67 citations counted in INSPIRE as of 11 Jul 2018
  
  
%\cite{Kallweit:2015dum}
\bibitem{Kallweit:2015dum} 
  S.~Kallweit, J.~M.~Lindert, P.~Maierhofer, S.~Pozzorini and M.~Schönherr,
  %``NLO QCD+EW predictions for V + jets including off-shell vector-boson decays and multijet merging,''
  JHEP {\bf 1604}, 021 (2016)
  doi:10.1007/JHEP04(2016)021
  [arXiv:1511.08692 [hep-ph]].
  %%CITATION = doi:10.1007/JHEP04(2016)021;%%
  %105 citations counted in INSPIRE as of 11 Jul 2018




%\cite{CMS:rwa}
\bibitem{CMS:rwa} 
  [CMS Collaboration],
  %``Search for new physics in monojet events in pp collisions at sqrt(s)= 8 TeV,''
  CMS-PAS-EXO-12-048.
  %%CITATION = CMS-PAS-EXO-12-048;%%
  %142 citations counted in INSPIRE as of 11 Jul 2018


  
  \bibitem{WZ_theory}
  %\cite{Lindert:2017olm}
%\bibitem{Lindert:2017olm} 
  J.~M.~Lindert {\it et al.},
  %``Precise predictions for $V+$ jets dark matter backgrounds,''
  Eur.\ Phys.\ J.\ C {\bf 77}, no. 12, 829 (2017).
 % doi:10.1140/epjc/s10052-017-5389-1
 % [arXiv:1705.04664 [hep-ph]].
  %%CITATION = doi:10.1140/epjc/s10052-017-5389-1;%%

\bibitem{monojet_exp}
%\cite{Aaboud:2017phn}
%\bibitem{Aaboud:2017phn} 
  M.~Aaboud {\it et al.} [ATLAS Collaboration],
  %``Search for dark matter and other new phenomena in events with an energetic jet and large missing transverse momentum using the ATLAS detector,''
  JHEP {\bf 1801}, 126 (2018).
%  doi:10.1007/JHEP01(2018)126
%  [arXiv:1711.03301 [hep-ex]].
  %%CITATION = doi:10.1007/JHEP01(2018)126;%%

\bibitem{Hooper}  
%\cite{Hooper:2013nia}
%\bibitem{Hooper:2013nia} 
See, for example, 
D.~Hooper,
%``Is the CMB Telling Us that Dark Matter is Weaker than Weakly Interacting?,''
Phys.\ Rev.\ D {\bf 88}, 083519 (2013), and references therein.
%doi:10.1103/PhysRevD.88.083519
%[arXiv:1307.0826 [hep-ph]].
%%CITATION = doi:10.1103/PhysRevD.88.083519;%%
%14 citations counted in INSPIRE as of 20 Apr 2018  


\end{thebibliography}
 \end{document}